\documentclass[manuscript,screen]{acmart}

\AtBeginDocument{%
  \providecommand\BibTeX{{%
    \normalfont B\kern-0.5em{\scshape i\kern-0.25em b}\kern-0.8em\TeX}}}

\copyrightyear{2026}
\acmYear{2026}
\setcopyright{cc}
\setcctype{by}
\acmConference[AutomotiveUI '26]{18th International Conference on Automotive User Interfaces and Interactive Vehicular Applications}{September 20--23, 2026}{Gothenburg, Sweden}
\acmBooktitle{18th International Conference on Automotive User Interfaces and Interactive Vehicular Applications (AutomotiveUI '26), September 20--23, 2026, Gothenburg, Sweden}
\acmDOI{10.1145/3828157.3828785}
\acmISBN{979-8-4007-2814-3/2026/09}

\usepackage{subcaption} 
\usepackage{url}
\usepackage{verbatim}
\makeatletter
\g@addto@macro{\UrlBreaks}{\UrlOrds}
\makeatother
\usepackage{multirow} %tabular cells spanning multiple rows
\usepackage{color} %colour management
\usepackage{enumitem} 

\usepackage{longtable}
\usepackage{todonotes}
\usepackage{xspace}

\usepackage{arydshln}

\newcommand\itema{\item[\textbf{\textcolor{blue}{RQ1}}]}
\newcommand\itemb{\item[\textbf{\textcolor{blue}{RQ2}}]}
\newcommand\itemc{\item[\textbf{\textcolor{blue}{RQ3}}]}

\newcommand{\m}{\textit{M=}}

\newcommand{\sd}{\textit{SD=}}

\newcommand{\N}{\textit{N=}}

\newcommand{\F}[3]{$F({#1},{#2})={#3}$}
\newcommand{\p}{\textit{p=}}
\newcommand{\padj}{\textit{adj. p=}}
\newcommand{\padjminor}{\textit{adj. p$<$}}

\newcommand{\pminor}{\textit{p$<$}}

\newcommand{\GroupID}{\textit{visualization condition}\xspace}
\newcommand{\optimization}{\textit{visualization condition}\xspace}

\newcommand{\trust}{\textit{trust}\xspace}
\newcommand{\predictability}{\textit{predictability}\xspace}

\newcommand{\perceivedSafetyScore}{\textit{perceived safety}\xspace}

\newcommand{\AcceptanceTwo}{\textit{satisfying}\xspace}

\newcommand{\longitudinalBo}{\textit{multi-session HITL MOBO}\xspace}
\newcommand{\run}{\textit{day}\xspace}

\newcommand{\opticar}{\textsc{OptiCarVis}\xspace}

\begin{document}

\title[Multi-Session User Experience Assessments of Computationally Optimized AV Functionality Visualizations]{Multi-Session User Experience Assessments of Computationally Optimized Automated Vehicle Functionality Visualizations}

\author{Mark Colley}
\authornote{Both authors contributed equally to this research.}
\email{m.colley@ucl.ac.uk}
\orcid{0000-0001-5207-5029}
\affiliation{%
  \institution{UCL Interaction Centre}
  \city{London}
  \country{United Kingdom}
}

\author{Pascal Jansen}
\authornotemark[1]
\email{pascal.jansen@uni-ulm.de}
\orcid{0000-0002-9335-5462}
\affiliation{%
  \institution{Institute of Media Informatics, Ulm University}
  \city{Ulm}
  \country{Germany}
}
\affiliation{%
  \institution{UCL Interaction Centre}
  \city{London}
  \country{United Kingdom}
}

\author{Svenja Krauß}
\email{svenja.krauss@uni-ulm.de}
\orcid{0009-0002-4047-0130}
\affiliation{%
  \institution{Institute of Media Informatics, Ulm University}
  \city{Ulm}
  \country{Germany}
}

\author{Enrico Rukzio}
\email{enrico.rukzio@uni-ulm.de}
\orcid{0000-0002-4213-2226}
\affiliation{%
  \institution{Institute of Media Informatics, Ulm University}
  \city{Ulm}
  \country{Germany}
}

\renewcommand{\shortauthors}{Colley \& Jansen et al.}

\begin{abstract}
Understanding automated vehicles (AVs) is crucial to improving their acceptance. Numerous approaches to visualizing relevant traffic information to passengers have been proposed and empirically evaluated. As this is time-consuming, costly, and reduces the possible design parameters, we employed multi-objective Bayesian optimization to optimize the design of visualizations in AVs. In particular, we evaluated multi-session aspects involving iterative optimization. We optimized the design for passenger trust and perceived safety while minimizing cognitive load. Results from an online study (N=74) show that this method effectively identifies visualization design parameter values that improve trust, safety, and predictability while making the design process more efficient and scalable. However, shortcomings of the computational approach when optimizing for subjective measurements are highlighted and discussed.
\end{abstract}




\begin{teaserfigure}
\centering
    \includegraphics[width=0.8\textwidth]{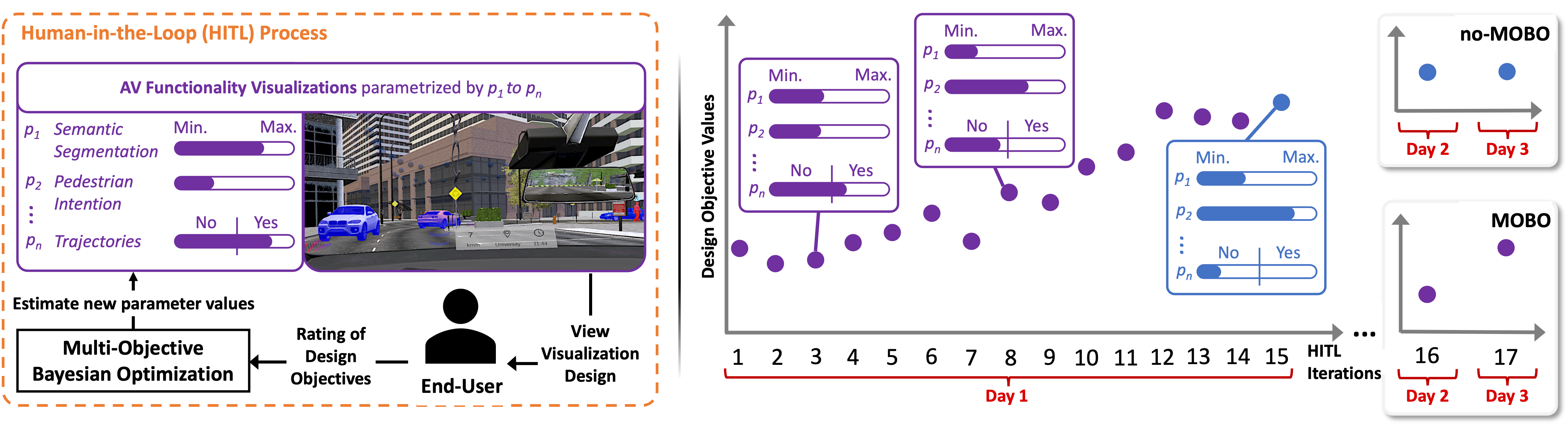}
  \caption{Human-in-the-loop (HITL) Multi-Objective Bayesian optimization (MOBO) of automated vehicle functionality visualization design to improve user ratings on \textit{multiple} design objectives like trust, safety, acceptance, and aesthetics while minimizing cognitive load. MOBO uses parameter values from the design and user ratings in each HITL iteration to suggest optimized design parameter values ($p_1$ to $p_n$) for the next one. In this \textbf{multi-session assessment}, participants used MOBO on day one to refine the design. On days two and three, the \textbf{MOBO} group further optimized their design, while the \textbf{no-MOBO} group viewed their original design.}
  \label{fig:teaser}
  \Description{Human-in-the-loop (HITL) multi-objective Bayesian optimization (MOBO) of automated vehicle functionality visualization design to improve user ratings on multiple design objectives like trust, safety, acceptance, and aesthetics while minimizing cognitive load. MOBO uses parameter values from the design and user ratings in each HITL iteration to suggest optimized parameter values for the next iteration. In this multi-session assessment, participants used MOBO on day one to refine the design. On days two and three, the MOBO group optimized their design, while the non-MOBO group viewed their original design.}
\end{teaserfigure}

\maketitle

\section{Introduction}

% Motivation
As automated vehicles (AVs) will fundamentally change mobility and traffic~\cite{fagnant2015preparing}, their exterior and interior designs must be adapted. Due to the possibility of engaging in non-driving-related-tasks such as reading or even sleeping~\cite{pfleging2016investigating,jansen2022design}, the focus shifts from driving safety and avoiding driver distraction to comfort and acceptance questions~\cite{colley2020effect, colley2021effects, colley2022scene, schneider2021explain}. 
Without acceptance, AVs might not be used. 
\citet{schoettle2014survey} report that 75\% were at least slightly concerned that AVs could fail in unexpected situations. This finding is supported by \citet{kyriakidis2015public}, who found that potential users worry about AVs' reliability. This might constitute undertrust, with acceptance and use likely to be scarce. However, overtrust, which means trusting the AV too much for its actual capabilities, could lead to dangerous situations. For example, users might completely disengage and fall asleep rather than monitor the AV. Therefore, previous work has suggested highlighting other road users in fog~\cite{winter2019assessing} or visualizing the inner workings of AVs, such as object detection~\cite{colley2021effects}, pedestrian intention prediction~\cite{colley2020effect}, or vehicle trajectory planning~\cite {colley2022scene}. %Additionally, for connected automated driving (CAD), \citet{mueller2022ar4cad} proposed visualizations of the area covered by external sensors (called the CAD-covered area) and of vehicles not perceivable by the AV without external support (called occluded cars). 

% Problem
While prior work~\cite{colley2022scene, mueller2022ar4cad, colley2020effect, colley2021effects} proposes fixed designs based on literature and intuition, such “one-size-fits-all” approaches are likely inadequate for the highly personal vehicular context. A key challenge is identifying suitable designs within a multi-dimensional space that balance competing objectives, such as perceived safety and cognitive load, without overwhelming the user.

Traditional design approaches used the user-centered design process~\cite{ISO9241_210_2019}, ISO standards (e.g.,~\cite{ISO15005_2017}), and guidelines such as the \textit{JAMA Guidelines for In-vehicle Display Systems}~\cite{JAMAGuideline2008}. These methods also relied on designer intuition, particularly when creating new AV experiences, often leading to time-consuming, resource-intensive user evaluations.

To align visualization designs with passengers' diverse needs, prior studies have explored personalization approaches that allow adjustments to icon size, location, and color. Yet, these studies typically do not explore continuous design parameter values (e.g., transparency of road user highlighting) and focus on a limited set of pre-defined options, which may fail to fully meet user preferences and overlook important design objectives besides usability, like perceived safety and trust~\cite{normark2015design, zhong2023evaluation, yunuo2022usability, adnan2018trust}.

Involving end-users in the design process through Human-in-the-Loop (HITL) Multi-Objective Bayesian Optimization (MOBO) can optimize design parameters based on subjective ratings, such as perceived safety and trust. However, this approach faces challenges due to the subjective nature of these metrics and the complexity of balancing multiple design objectives, which can result in inconsistencies and disregard users' prior knowledge and preferences~\cite{chan2022bo, liao2023interaction, chandramouli2023mobopersonalize, koyama2022boassistant, kadner2021adaptifont, ou2022infite}.
Therefore, Jansen \& Colley et al. \cite{jansen2025opticarvis} introduced \opticar, a method for optimizing the design of AV functionality visualization through HITL MOBO. Their approach aimed to enhance end-user perceptions of safety and trust, improve understanding of AV operations, minimize cognitive load, and enhance the perceived usefulness, satisfaction, and visual appeal of the visualizations. They engaged end-users with non-technical or non-design backgrounds, leveraging their insights, experiences, and preferences in the HITL optimization process.
In their demonstration of \opticar, they designed visualizations displayed on a Head-Up Display (HUD), including the AV's functional levels (i.e., \textit{Situation Detection}, \textit{Situation Prediction}, and \textit{Trajectory Planning}), the connected automated driving (CAD)-covered area and occluded cars, and general information such as AV speed, destination, and current time, to convey relevant information at various stages of the automated driving task~\cite{jansen2025opticarvis}.

% Position your own approach
We built on this work to evaluate the \textbf{multi-session} effects of \opticar over three days (see \autoref{fig:teaser}).
%with direct user involvement and a comparison to a baseline without visualization and a baseline visualization inspired by the previous work~\cite{kunze2018augmented, colley2020effect, colley2022scene, mueller2022ar4cad}. 
%% implementation 
%We optimize AV visualizations using visual elements like a functional hierarchy, road layout, and real-time status on a HUD, informed by existing research~\cite{colley2022scene,mueller2022ar4cad,flohr2023prototyping}. These visuals incorporate elements like semantic segmentation and pedestrian intent icons, parameterized by transparency and size. Position parameters were held constant based on prior work~\cite{colley2022scene,mueller2022ar4cad}.
%Our Unity simulation allows users to experience these through an AR Windshield Display. During the Human-in-the-Loop (HITL) process, a Multi-Objective Bayesian Optimization (MOBO) algorithm fine-tunes these parameters. This Unity app bridges user feedback with MOBO for iterative improvements.
In line with Jansen \& Colley et al. \cite{jansen2025opticarvis}, we explored five design strategies (C2-C6) and, additionally, no visualization:

\indent C1-\textit{No Visualization}.\newline
\indent C2-\textit{Custom design by experts}, showing averaged design parameter values based on \N8 expert designs.\newline
\indent C3-\textit{Custom design by end-users}, where they could use a design parameter tool to adjust the values.\newline
\indent C4-\textit{Cold-Start HITL MOBO} with random initial settings.\newline
\indent C5-\textit{Expert-Informed Warm-Start HITL MOBO} with the expert design as the initial setting.\newline
\indent C6-\textit{User-Informed Warm-Start HITL MOBO} with users' custom design as initial setting.
\smallskip

%% multi-session method
After generating their final design on day one in C2-C6, participants were randomly assigned to a \longitudinalBo group: either \textbf{(1)} the HITL MOBO group, where the MOBO remained active or started based on the participant feedback on day one, or \textbf{(2)} the group without HITL MOBO, where participants experienced their final design of day one over the remaining days. No visualization was shown for C1 over the three days.

In a between-subject study with \N74, we found that \longitudinalBo enhanced \textbf{all} performance-related aspects of the visualization. 
These were cognitive load~\cite{hart1988development}, trust~\cite{korber2018theoretical}, predictability~\cite{korber2018theoretical}, perceived safety~\cite{faas2020longitudinal}, aesthetics, and acceptance~\cite{van1997simple}. This is especially interesting as these values were close \emph{across all conditions and both multi-session groups} on day one.

These findings show that a multi-session approach to optimizing AV functionality visualizations can maintain and improve users’ subjective perceptions over multiple days, even with infrequent feedback. In practice, this argues for an initial familiarization and personalization phase that may use HITL MOBO to enable rapid adaptation, after which periodic re-optimization can account for changing user perceptions without requiring frequent updates or feedback.

\smallskip
\noindent \textbf{Contribution Statement:}
This work (1) contributes a three-day multi-session empirical investigation (\N74) of optimization-driven design approaches in the automotive visualization context.
Additionally, we provide (2) source code for the driving environment, the evaluation scripts, and the anonymized data.

\section{Background and Related Work}
This work builds on research in automotive visualization design and computationally optimized User Interfaces (UIs).
% to-do cite: Evaluation and Optimization of In-Vehicle HUD Design by Applying an Entropy Weight-VIKOR Hybrid Method - https://www.mdpi.com/2076-3417/13/6/3789

% to-do cite: https://dl.acm.org/doi/10.1145/3565472.3592956

\subsection{In-Vehicle Visualizations of Automated Vehicle Functionalities}

%Different methods of communicating driving aspects, such as decisions, detections, destination, regulation, and navigation, have been assessed in prior research. Ambient light has been used to convey AV decisions to users, as demonstrated by \citet{locken2016autoambicar}. Similarly, \citet{wilbrink2020reflecting} suggested using light strips to indicate intention or perception. \citet{lindemann2018catch} used an Augmented Reality (AR) Windshield Display (WSD) to emphasize potential threats like pedestrians and utilized a cube over moving vehicles to denote their behavior. This led to enhanced situational awareness, beyond just having fundamental elements like speed and navigation information.

Prior studies have assessed various display technologies, including HUDs, LED strips, and Augmented Reality (AR) Windshield Displays (WSDs), for presenting diverse driving-related information in AVs. %For example, \citet{colley2020effect} discovered that the AR WSD effectively decreased cognitive load.

%The concept of \textit{Calibrated trust}\cite{muir1996trust} is defined as a state where a user's trust level aligns with the automated system's capabilities, thereby avoiding issues of over- and undertrust. 

For example, \citet{hauslschmid2017supportingtrust} demonstrated that visualizing an AV's current interpretation through a miniature world or a simulated avatar can enhance trust, although the reported need for such visualizations varied among participants. Similarly, \citet{currano2021little} found that an AR HUD improved situational awareness depending on the scene complexity and driver behavior, indicating a need for personalization.
\citet{schneider2021explain} showed that AR WSDs and LED strips that provide explanations of the AV's future trajectory enhance the user experience. However, supplementary explanations via smartphone apps did not add value.
\citet{colley2021should} found that an abstract HUD representation (a symbol with text, e.g., for crossing animals) was sufficient to convey critical information, as reflected in participants' ratings.

Highlighting the role of uncertainty visualization, i.e., the reliability of the AV in performing, for example, object detection. \citet{beller2013improving} found a simple anthropomorphic uncertainty symbol increased situational awareness and trust, whereas \citet{helldin2013presenting}, using abstract bars, found users took control sooner but trusted the automation \emph{less} when uncertainty was shown---displaying uncertainty thus raised trust in one case and lowered it in the other.
Whereas the above are dashboard symbols, AR overlays can situate uncertainty in the scene: \citet{kunze2018augmented} found hue particularly effective for conveying urgency. \citet{colley2021effects} also pointed out that such abstract visualizations might obscure the source of uncertainty. They used semantic segmentation in AR to enhance situational awareness without altering trust or cognitive load.

Exploring further, \citet{colley2022scene} compared different AV visualization levels and combinations, finding that displaying the planned ego trajectory increased trust, while visualizing the \emph{predicted trajectories of other road users} increased cognitive load. \citet{flohr2023prototyping} and \citet{10.1145/3610886} confirmed the importance of visualizing AV functionalities (situation detection and the planned trajectory) for maintaining appropriate levels of trust, showing that such visualizations could predictably enhance usefulness and user experience in studies conducted in real vehicles.
%Lastly, \citet{mueller2022ar4cad} reused several visualization designs, such as connectivity symbols and planned trajectories, to depict occluded vehicles and facilitate merging, ultimately achieving higher levels of trust, reliability, and understanding by integrating multiple visualizations within the CAD context.

%\citet{mackay2020impact}

\subsection{Personalization and Computational Methods of In-Vehicle Interface Design}\label{rw-optimization-auto-ui}

Recent studies have demonstrated the advantages of personalized in-vehicle interfaces, emphasizing safety, trust, and acceptance~\cite{ayoub2019tenyears,micklitz2023design}. Manual customization by end-users, such as adjustments explored by Normark (e.g., icons' size, location, and color on the dashboard), enhances usability~\cite{normark2015design}. However, manual customization can introduce safety-relevant errors (e.g., overlapping or low-contrast elements); because even experts can misjudge these effects, designs need to be evaluated with users rather than by expertise alone. In contrast, computational methods, as discussed by Zhong and Yunuo, utilize ratings to optimize design. However, these methods often overlook iterative user feedback and broader design possibilities~\cite{zhong2023evaluation,yunuo2022usability}.

To address these gaps, computational optimization methods, incorporating iterative refinement and multi-objective considerations, are being explored for more effective personalization~\cite{chan2022bo}. This research employs HITL optimization, integrating designer and user inputs to better align visualization design with end-user preferences while balancing multiple objectives, such as safety and cognitive load.

\subsection{Human-in-the-Loop Optimization and Multi-objective Optimization}

HITL optimization incorporates human feedback in iterative cycles for parameter tuning, particularly for design tasks requiring subjective evaluations~\cite{chiu2020human, koyama2020sequential, takagi2001interactive, zhong2021spacewalker}. Bayesian Optimization (BO) is a machine learning approach for optimizing complex functions and is well-suited for HITL processes~\cite{chan2022bo, dudley2019crowdsourcing, khajah2016games}. It excels in handling noisy data and efficiently sampling the design space.
HITL and BO have been applied to various design problems, such as interactive menu optimization~\cite{bailly2013menuoptimizer} and wearable device tuning using physiological measures~\cite{kim2017human}. These methods enable experts and end-users to actively contribute to the design process. However, traditional HITL and BO approaches often focus on single objectives, which may not suffice for complex systems, such as in-vehicle visualizations in AVs, that require balancing multiple objectives, such as safety and user trust.
MOBO addresses multiple design objectives simultaneously by generating a Pareto front to represent trade-offs~\cite{dunlop2012multidimensional, dridhar2015midair, hayward1994design, chandramouli2023mobopersonalize}. Given its demonstrated efficacy in HCI and based on recent work on HITL MOBO in the automotive UI domain \cite{jansen2025opticarvis,colley2025improving}, we posit that MOBO is suitable for optimizing in-vehicle visualizations.

\subsection{Comparing Designer and Human-In-The-Loop Optimization}

\citet{chan2022bo} and \citet{liao2023interaction} found that designers felt disconnected from MOBO-driven designs, despite their quality. \citet{koyama2022boassistant} bridged this gap by using BO as a design assistant, allowing designers to blend their expertise with BO suggestions.
However, for in-vehicle visualizations, end-users' insights are crucial, as they possess unique preferences and perceptions. While prior research focused on designer integration in HITL MOBO processes, end-user involvement remains underexplored. This work emphasizes engaging end-users in in-vehicle visualization design to overcome existing HITL MOBO challenges. We also expand the understanding of HITL optimization qualities by contrasting end-user-led and optimizer-driven processes~\cite{chan2022bo,liao2023interaction}. Finally, the most important distinguishing factor compared to all previous work on HITL MOBO is our focus on multi-session effects, which are underexplored. %, especially when the HITL MOBO is applied every day. 

\subsection*{Research Gap}\label{rw-personalized-auto-ui}

Despite a decade of research highlighting the benefits of personalization in the automotive domain~\cite{ayoub2019tenyears,micklitz2023design}, there remains a gap in methods for their optimization. Manual personalization can be user-friendly, but technically challenging and error-prone. Computational optimization methods, on the other hand, offer a more reliable and adaptable approach to design, particularly when considering user preferences~\cite{chan2022bo}.

Existing computational strategies, such as those by \citet{zhong2023evaluation} and \citet{yunuo2022usability}, optimize for specific objectives but lack an inclusive approach that accounts for designer and end-user input. These methods may also leave large parts of the (effectively unbounded) design space unexplored, potentially overlooking better solutions.
To overcome these challenges, approaches emerged using HITL MOBO to adapt designs to users with diverse preferences, abilities, and needs. For instance, optimizing the design of AV functionality visualizations \cite{jansen2025opticarvis}, external communication via exterior LED strips with pedestrians to indicate safe road crossing \cite{colley2025improving}, and functionality visualizations of air taxis (e.g., flight path and detected other air taxis) in the context of urban air mobility \cite{meinhardt2025fly}.

Our work distinguishes itself by employing a HITL MOBO approach that integrates multi-session insights from designers and end-users. We optimize automotive functionality visualizations across multiple subjective objectives,safety, trust, and acceptance, \textbf{across multiple sessions}.

\section{Optimization of Automated Vehicle Functionality Visualization}
\begin{figure*}[ht]
    \centering
    \includegraphics[width=0.8\linewidth]{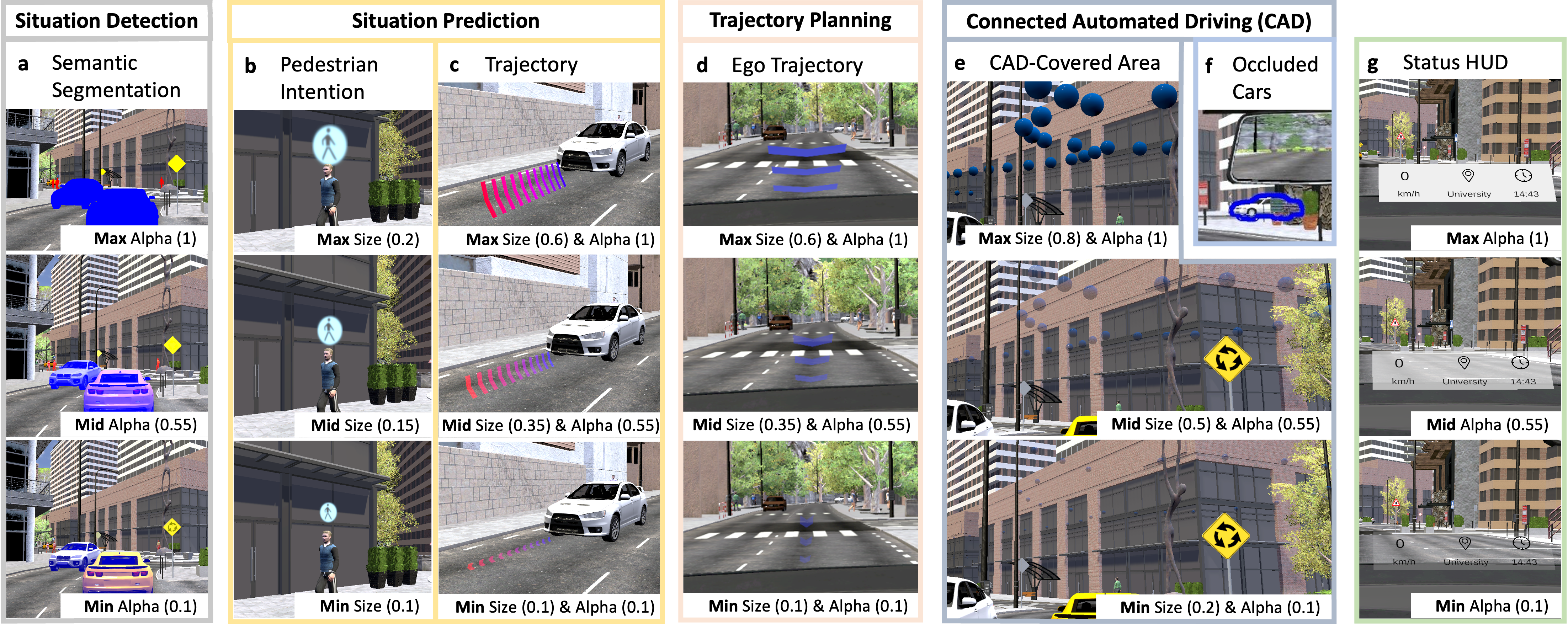}
    \caption{Overview of the employed visualizations of an SAE Level 4 \cite{SAElevel} AV's functional levels of internal operation \cite{colley2022scene}, CAD \cite{mueller2022ar4cad}, and status on an AR WSD, showing the possible variations in transparency (alpha) and size values (see brackets). \textit{Min} and \textit{Max} represent the designs at the lower and upper bounds of the continuous parameter ranges, while \textit{Mid} represents the midpoints (see \cite{jansen2025opticarvis}).}
    \label{fig:in-vehicle-visualizations-overview}
    \Description{The figure provides an overview of seven visualization concepts, showcasing them in action. For each concept, three variations display different levels of transparency and size. Shown are the trajectory, pedestrian intention, semantic segmentation, occlusion highlighting, ego trajectory, CAD-covered area, and car status in the HUD.}
\end{figure*}

The AV functionality visualization design is derived from previous work regarding its effects on trust, cognitive load, and perceived safety \cite{jansen2024visualizing}. Specifically, we built upon the work of \citet{colley2022scene}, who introduced the idea of visualizing the functionalities of AVs: ''Situation Detection'', ''Situation Prediction'', and ''Trajectory Planning''. This includes the visualization of the object detection and its subsequent colorization~\cite{colley2021effects}, the depiction of pedestrian intention as a symbol~\cite{colley2020effect}, the inferred future trajectory of other vehicles~\cite{colley2022scene}, and the own planned trajectory~\cite{colley2022scene}. Additionally, we add visualizations from \citet{mueller2022ar4cad}, who introduced visualizations for CAD. \autoref{fig:in-vehicle-visualizations-overview} provides an overview of the design space.

\subsection{Bayesian Optimization: Design Parameters, Design Objectives, and Setup}\label{sec:bo}

%\subsubsection{Design Parameters, Bayesian Optimizer, Objective Functions, Hyperparameter Setup, and Stop Criterion}\label{study-apparatus-bo}
Our work builds directly on the work by Jansen \& Colley et al. \cite{jansen2025opticarvis}. We employed the same visualization design parameter in the same ranges, summarized in \autoref{tab:design_param} and shown in \autoref{fig:in-vehicle-visualizations-overview}. We also used the same design objectives measured via subjective metrics: cognitive load (via NASA-TLX~\cite{hart1988development}), trust and understandability \cite{korber2018theoretical}, perceived safety~\cite{faas2020longitudinal}, acceptance~\cite{van1997simple}, and visual appeal~\cite{colley2023uam}. Free-text feedback that participants could give was \emph{not} fed back into the MOBO. Besides, we replicated the MOBO hyperparameters, HITL iteration setup (i.e., 20 iterations on day one), and HITL process stop criterion (i.e., whether participants give the perfect rating for \textbf{every} subjective metric) as in \cite{jansen2025opticarvis}. The MOBO implementation used \href{https://botorch.org/}{\texttt{BoTorch}} in version 0.9.2.

\begin{table*}[ht!]
\scriptsize
\caption{The 16 design parameters for the visualization design, with the ranges. All design parameters are modeled continuously, with values mapped to Boolean if necessary (''Bool''), see \cite{jansen2025opticarvis}.}
\label{tab:design_param}
%\resizebox{\textwidth}{!}{%
\begin{tabular}{@{}llll@{}}
\toprule
\textbf{Design Parameter }                  & \textbf{Description}                                  & \textbf{Reference}                 & \textbf{Range}                  \\ \midrule
$x_1$: Semantic Segmentation, $v$       & Whether the semantic segmentation result should be visualized.   &~\cite{colley2021effects}  & [0, 1]; Bool \\
$x_2$: Semantic Segmentation Alpha, $\alpha$ & Alpha value of the semantic segmentation.                        &~\cite{colley2021effects}  & [0.1, 1]             \\ \hdashline

$x_3$: Pedestrian Intention, $v$        & Whether the predicted pedestrian intention should be visualized. &~\cite{colley2020effect}   & [0, 1]; Bool \\
$x_4$: Pedestrian Intention Size, $s$   & Alpha value of the pedestrian intention symbol.                  &~\cite{colley2020effect}   & [0.1, 0.2]             \\ \hdashline

$x_5$: Trajectory, $v$                  & Whether the predicted trajectory of others should be visualized. &~\cite{kunze2018augmented} & [0, 1]; Bool \\ 
$x_6$: Trajectory Alpha, $\alpha$            & Alpha value of the trajectory.                                   &~\cite{kunze2018augmented} & [0.1, 1]             \\
$x_7$: Trajectory Size, $s$             & Size of the trajectory.                                          &~\cite{kunze2018augmented} & [0.1, 0.6]             \\ \hdashline

$x_8$: Ego Trajectory, $v$   & Whether the own planned trajectory should be visualized.                  &~\cite{colley2020effect}   &  [0, 1]; Bool         \\ 
$x_{9}$: Ego Trajectory Alpha, $\alpha$   & Alpha value of the own planned trajectory.                  &~\cite{colley2020effect}   & [0.1, 1]              \\ 
$x_{10}$: Ego Trajectory Size, $s$             & Size of the own planned trajectory.                                          &~\cite{kunze2018augmented} & [0.1, 0.6]             \\ \hdashline

$x_{11}$: CAD-Covered Area, $v$                & Whether the area covered through V2x should be visualized.       &~\cite{mueller2022ar4cad}  & [0, 1]; Bool \\
$x_{12}$: CAD-Covered Area Alpha, $\alpha$         & Alpha value of the symbols for the CAD-covered area.                     &~\cite{mueller2022ar4cad}  & [0.1, 1]    \\    
$x_{13}$: CAD-Covered Area Size, $s$         & Size of the symbols for the CAD-covered area.                     &~\cite{mueller2022ar4cad}  & [0.2, 0.8]    \\ \hdashline

$x_{14}$: Occluded Cars, $v$              & Whether occluded (e.g., by buildings) cars should be visualized. &~\cite{mueller2022ar4cad}  & [0, 1]; Bool \\ \hdashline

$x_{15}$: Vehicle Status HUD, $v$              & Whether the vehicle status in the HUD should be visualized.      &~\cite{currano2021little}                          & [0, 1]; Bool \\
$x_{16}$: Vehicle Status HUD Alpha, $\alpha$            & Alpha value of the vehicle status.       &~\cite{riegler2019adaptive}                & [0.1, 1]\\ \bottomrule
\end{tabular}%
%}
\end{table*}

\subsection{Design and Optimization Conditions}\label{study-optimization-strategies}

Besides an option with no visualization, we employed the conditions of Jansen \& Colley et al. \cite{jansen2025opticarvis} (hereafter called \GroupID) for participants' initial visualization designs on day one:

\paragraph{C1 No Visualization}\label{c1}
In this condition, no AV functionality is displayed.
\vspace{-0.3cm}
\paragraph{C2 Custom design by experts}\label{s4}
Instead of HITL optimization, end-users assess a \textit{standard} visualization design of automotive UI experts (\N8) using the parameter design tool. This design uses the average parameter values of the experts.
\vspace{-0.3cm}
\paragraph{C3 Custom design by end-users}\label{s5}
Instead of an expert-crafted \textit{standard} design (see C2), end-users create and personalize their own visualizations using the parameter design tool. These custom designs are then evaluated after the AV ride.
\vspace{-0.3cm}
\paragraph{C4 Cold-start HITL MOBO}\label{s1}
The \textit{cold-start} HITL MOBO starts with randomly selected parameters from the Bayesian optimizer. End-users subsequently engage with and rate prospective designs, which is the feedback used by the optimizer to refine the design parameter values.
\vspace{-0.3cm}
\paragraph{C5 Expert-Informed warm-start HITL MOBO}\label{s2}
In the \textit{warm-start} variant, automotive UI professionals (\N8) used a design tool to scrutinize and select design parameters before initializing the HITL MOBO to narrow the design space. % in line with industry best practices.
\vspace{-0.3cm}
\paragraph{C6 User-Informed warm-start HITL MOBO}\label{s3}
End-users explore and choose initial visualization designs, kickstarting a \textit{warm-start} HITL MOBO. Like C5, this method aims to expedite the optimizer's discovery of optimal designs by leveraging the users' prior knowledge and preferences~\cite{chan2022bo,liao2023interaction}. This adaptation could also alleviate feelings of low agency associated with HITL methods~\cite{chan2022bo}.

\paragraph{Multi-Session Optimization}\label{s_long}
After the designs for participants were generated (C4-C6), self-defined (C3), or set by the expert design (C2) on day one, for the multi-session study, participants were randomly selected to either be in a group without further MOBO over the three days or whether the MOBO would be again started with the results of the assessments of the previous day.

\section{Experiment}

We investigate how, in a \textbf{multi-session} setting, in-vehicle visualizations defined by the MOBO affect potential users:
\begin{itemize}
    \itema How are end-users' perceptions (safety, trust, predictability, usefulness, satisfaction, aesthetics) and cognitive load affected over three sessions, and does keeping the HITL MOBO active---versus freezing the day-one design---change this trajectory?
    \itemb Which day-one design strategy (C2--C6) yields the highest ratings, and do the warm-start strategies (C5/C6) differ from the cold-start (C4) or from the static custom designs (C2/C3)?
    \itemc How do the optimization dynamics evolve across sessions (convergence, day-to-day design drift, and between-user personalization)?
\end{itemize}

The six conditions isolate the two factors these questions require. C1 (No Visualization) is a baseline testing whether \emph{any} functionality visualization helps. C2--C3 represent the current state of the art of \emph{static} design (an averaged expert design and a self-made end-user design), while C4--C6 are the three HITL MOBO variants (cold-, expert-warm-, and user-warm-start). Crossing C2--C6 with the multi-session factor (HITL MOBO kept active vs.\ design frozen after day one) separates the effect of \emph{having} an optimized design from the effect of \emph{continuing} to optimize it---the central question of a multi-session deployment.

The experimental procedure followed the ethics committee guidelines of our university and adhered to regulations on handling sensitive and private data, anonymization, compensation, and risk aversion. Compliant with our university‘s local regulations, no additional formal ethics approval was required.

\subsection{Apparatus}

%\subsubsection{Automated Vehicle and Driving Environment}
\label{study-apparatus-unity}
We developed a cross-platform application compatible with Windows and macOS, utilizing \href{https://unity.com/}{Unity} 2022.3.7. The application replicates in-vehicle visual displays within an automated driving context, featuring a 3D Tesla Model X modified with a virtual AR WSD and a vehicle status HUD; only the WSD/HUD overlays were study-controlled, the vehicle model itself was unchanged.
We employed the Unity Windridge City environment for the simulation, in line with existing studies~\cite{colley2020effect, colley2021effects, colley2022scene}. We integrated the \href{https://assetstore.unity.com/packages/templates/systems/urban-traffic-system-89133}{Urban Traffic System} asset to model realistic traffic and pedestrian activities. The simulated AV follows a predetermined 33-second route for the MOBO runs, engineered to yield frequent interactions between pedestrians and vehicles, thereby creating various visualization scenarios.

The \textit{automotive UI experts} (N=8; 2 female, 6 male) specialized in in-vehicle UI usability and trust in automation, with backgrounds in psychology, computer science/HCI, and engineering across four institutions in Europe, the USA, and Canada. They are research associates, Ph.D.\ students, or (former) engineers at two large European OEMs, on average \m{27.88} (\sd{2.36}) years old, and have each published multiple papers on automotive design. Following \citet{shanteau2003can}, this combination of \emph{Experience} and \emph{Peer Identification} qualifies them as domain experts; the full expert procedure and the resulting averaged design are reported in~\cite{jansen2025opticarvis}.

\subsection{Participants}

\N74 US participants (Mean age = 38.5, SD = 12.2, range: [19, 72]; Gender: 32.4\% women, 66.2\% men, 1.35\% non-binary; Education: College, 71.62\%; High School, 22.97\%; Vocational training, 5.41\%) took part. 
%The participant pool was limited to US residents.
Regarding their employment status, 50 are employees; 6 are college students; 1 is at a school; 9 are self-employed; 5 are job-seeking; and 3 indicated \textit{other}.
All participants hold a valid driver's license for \m{18.30} (\sd{12.32}) years. 
We found no significant differences between the \optimization for license, gender, or age.
All volunteers were informed of consent and agreed to the recording and anonymized publication of results.
Participants were compensated with £3.80. % (£ due to Prolific based in the UK).

\subsection{Procedure}\label{sec:procedure_study}

We conducted the research through an online platform to better involve a variety of end-users, particularly those without technical expertise—something that often proves difficult in lab-based studies. Furthermore, using a Unity application to simulate AVs and present videos to participants safely is a commonly accepted approach for assessing new in-vehicle visualizations. This is particularly relevant as the AV technology required for a more direct assessment is inaccessible (e.g., see~\cite{jansen2022design,colley2020effect,colley2022scene}).

\begin{figure*}[ht!]
\centering
    \includegraphics[width=0.75\linewidth]{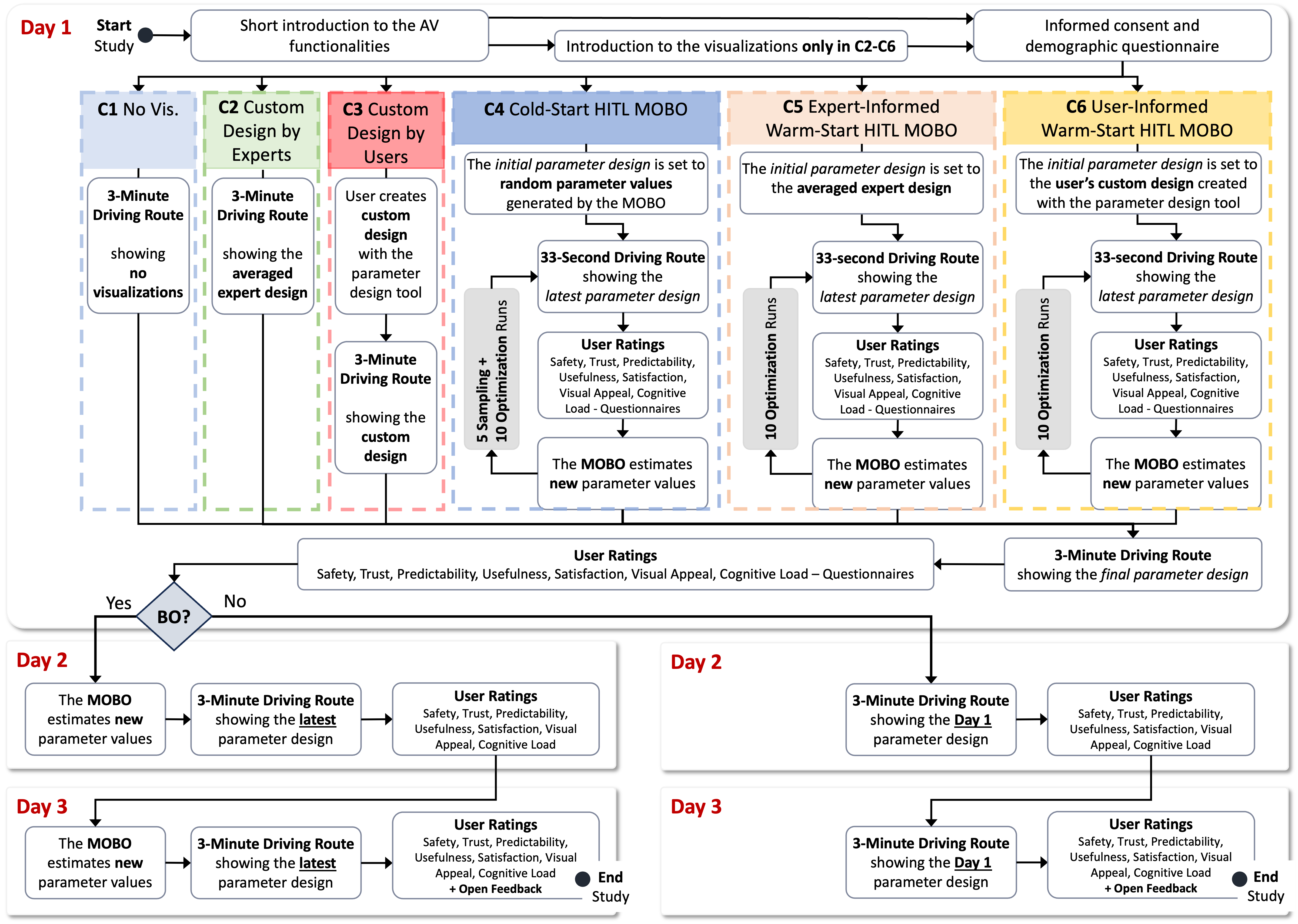}
   \caption{Study procedure shown before the multi-session part started for the different \optimization (see \cite{jansen2025opticarvis}). Participants were then randomly distributed (''BO?'') to \longitudinalBo groups.}~\label{fig:process}
    \Description{This figure shows the procedure for each of the six study groups, C1 - C6, using flow charts over three days.}
\end{figure*}

\paragraph{Day one.}
Day one used the single-session \opticar procedure and results~\cite{jansen2025opticarvis}. After the introduction (Appendix \autoref{fig:intro}), participants in the MOBO conditions (C4--C6) experienced the 33-second route once per HITL iteration---15 iterations for the cold-start C4 (5 sampling + 10 optimization) and 10 for the warm-starts C5/C6 (optimization only, as the initial design replaces the sampling phase)---rating the visualization after \emph{each} iteration. In C3 and C6, participants first created a design with the parameter-design tool. All participants then experienced a previously unseen 3-minute route and rated their final day-one design; for the static conditions C2 (averaged expert design) and C3 (own design), this 3-minute drive was the only ride. C1 (No Visualization) saw the routes without any overlay.

\paragraph{Days two and three.}
After day one, participants in C2--C6 were randomly assigned to one of two \textbf{multi-session groups}: \emph{HITL MOBO}, in which the optimizer ran one further round each day starting from the previous day's ratings, or \emph{no optimization}, in which the day-one design was frozen. On each of days two and three, participants experienced the 3-minute route once and rated the (possibly updated) design. C1 saw no visualization on all three days.

%Each session started with a brief introduction, informed consent, and a demographic questionnaire.  We instructed participants about the available visualizations (see Appendix \autoref{fig:intro}) and parameters (see \autoref{tab:design_param}). The introduction to the AV capabilities was taken from \citet{colley2022scene} (see Appendix Section \ref{appendix:participant-instructions}).  Each participant was randomly assigned to one of six conditions. After the first day, they were further divided into two \textbf{multi-session groups}: (1) the group using HITL MOBO again and (2) the group without it (i.e., \textit{no optimization}). Participants evaluated the final design over three consecutive days, with each session lasting about 8 minutes, for a total of 25 minutes. On the second and third days, participants again experienced a 3-minute drive.  This repeated exposure over three days aligns with prior multi-session studies investigating user perceptions of AVs \cite{colley2022repeated, colley2024longitudinal}. The HITL MOBO group used MOBO for one more round of optimization each day, continuing from where it left off the previous day, using participants’ previous ratings. In contrast, the \textit{no optimization} group rated the ride without the MOBO making any further changes to the design from the first day onward.
%Participants were compensated with €.

\subsection{Measurements}

\begin{table}[ht]
\caption{Subjective measures, items, scale, and source.}
\label{tab:measures}\scriptsize
\begin{tabular}{@{}lp{3.2cm}ll@{}}
\toprule
\textbf{Label} & \textbf{Item / construct} & \textbf{Scale} & \textbf{Src}\\\midrule
Cognitive load & NASA-TLX mental workload & 1--20 & \cite{hart1988development}\\
Trust & 2 items & 1--5 & \cite{korber2018theoretical}\\
Predictability & 4 items (understandability) & 1--5 & \cite{korber2018theoretical}\\
Perceived safety & 4 semantic differentials & $-3$/$+3$ & \cite{faas2020longitudinal}\\
Usefulness & ``\dots useful'' & 1--7 & \cite{van1997simple}\\
Satisfying & ``\dots satisfying'' & 1--7 & \cite{van1997simple}\\
Aesthetics & ``\dots visually appealing'' & 1--7 & \cite{colley2023uam}\\\midrule
Expectation & matches my imagination & 1--7 & \cite{chan2022bo}\\
Satisfaction & pleased with final \emph{design} & 1--7 & \cite{chan2022bo}\\
Confidence & design is optimal for me & 1--7 & \cite{chan2022bo}\\
Agency & in control of design process & 1--7 & \cite{chan2022bo}\\
Ownership & final design is mine & 1--7 & \cite{chan2022bo}\\\bottomrule
\end{tabular}
\end{table}

%\paragraph{Subjective Measurements}
In addition to the measurements used for the MOBO (see Section~\ref{sec:bo} and \autoref{tab:measures}), participants could provide textual feedback after each day on the following aspects. We measured design experience using questions adapted from \citet{chan2022bo}. On 7-point Likert scales (1=\textit{Strongly disagree} to 7=\textit{Strongly agree}), we queried about \textit{Expectation}: "The final design matches my imagination.", \textit{Satisfaction}: "I'm pleased with the final design.", \textit{Confidence}: "I believe the design is optimal for me.", \textit{Agency}: "I felt in control of the design process." and \textit{Ownership}: "I feel the final design is mine." %

% \paragraph{Objective Measurements}
% We integrated a webcam-based eye-tracker, UnitEye\footnote{To be open-sourced -- own development}, into our study application to monitor Areas of Interest (AOIs), including pedestrians, vehicles, traffic signs, pedestrian intent icons, occluded vehicles, CAD-covered areas, and the vehicle status HUD. Prior to the study, participants underwent eye-tracking calibration. This allowed us to observe where they were looking and what captured their attention during both the design phase and the autonomous vehicle (AV) ride. It should be noted, however, that this eye-tracking data did not influence the objective function used by the Bayesian optimizer.

\section{Results}

%\subsection{Data Analysis}
Before each statistical test, we checked the required assumptions (e.g., normality).
For non-parametric data, we used the ARTool package~\cite{wobbrock2011art}, as ANOVA is inappropriate for non-normally distributed data. The procedure is abbreviated, as in the original publication, with ART.
For pairwise comparisons, we used ART contrasts (\texttt{art.con} via estimated marginal means on the aligned-rank model) with Holm correction, which—unlike a between-groups rank test—respect the within-subject (repeated-day) structure. Because the no-visualization baseline (C1) has no \longitudinalBo counterpart, condition contrasts are computed over C2--C6 (C1 comparisons are reported in the day-one analysis~\cite{jansen2025opticarvis}). $p$-values are Holm-corrected within each objective. As a robustness check, we additionally fit linear mixed models (\texttt{lmerTest}, participant as random intercept), which reproduced every \longitudinalBo effect reported below. Only participants with complete three-day records entered the per-session models.
We employed R in version 4.6.0 and RStudio in version 2026.05.1. All packages were up to date in June 2026.

\begin{table*}[ht!]
\caption{Consolidated ART omnibus results for the seven performance objectives: $F$ (with degrees of freedom) for the main effects of \GroupID, \longitudinalBo, \run, and the \longitudinalBo$\times$\run interaction. Linear mixed models (participant random intercept) reproduce every \longitudinalBo effect. $^{*}p<.05$, $^{**}p<.01$, $^{***}p<.001$. Aesthetics, usefulness, and satisfying were collected only in the design-process conditions (hence reduced df).}
\label{tab:omnibus}\footnotesize
\begin{tabular}{@{}lcccc@{}}
\toprule
\textbf{Objective} & \textbf{\GroupID} & \textbf{\longitudinalBo} & \textbf{\run} & \textbf{\longitudinalBo$\times$\run} \\
\midrule
Cognitive load   & $F(5,65){=}0.46$       & $F(1,65){=}13.47^{***}$ & $F(2,130){=}9.24^{***}$ & $F(2,130){=}3.98^{*}$   \\
Trust            & $F(5,65){=}3.41^{**}$  & $F(1,65){=}23.42^{***}$ & $F(2,130){=}0.33$       & $F(2,130){=}2.56$       \\
Predictability   & $F(5,65){=}3.13^{*}$   & $F(1,65){=}24.26^{***}$ & $F(2,130){=}2.99$       & $F(2,130){=}2.78$       \\
Perceived safety & $F(5,65){=}2.95^{*}$   & $F(1,65){=}20.88^{***}$ & $F(2,130){=}1.09$       & $F(2,130){=}3.19^{*}$   \\
Aesthetics       & $F(4,50){=}2.77^{*}$   & $F(1,50){=}12.92^{***}$ & $F(2,100){=}2.52$       & $F(2,100){=}7.66^{***}$ \\
Usefulness       & $F(4,50){=}1.54$       & $F(1,50){=}14.36^{***}$ & $F(2,100){=}0.01$       & $F(2,100){=}6.67^{**}$  \\
Satisfying       & $F(4,50){=}4.51^{**}$  & $F(1,50){=}18.41^{***}$ & $F(2,100){=}0.51$       & $F(2,100){=}3.76^{*}$   \\
\bottomrule
\end{tabular}
\end{table*}

\subsection{Design Performance}

\paragraph{Cognitive Load}

\begin{figure*}[ht!]
\centering
    \begin{minipage}[t]{0.48\textwidth}
        \centering
        \includegraphics[width=0.8\linewidth]{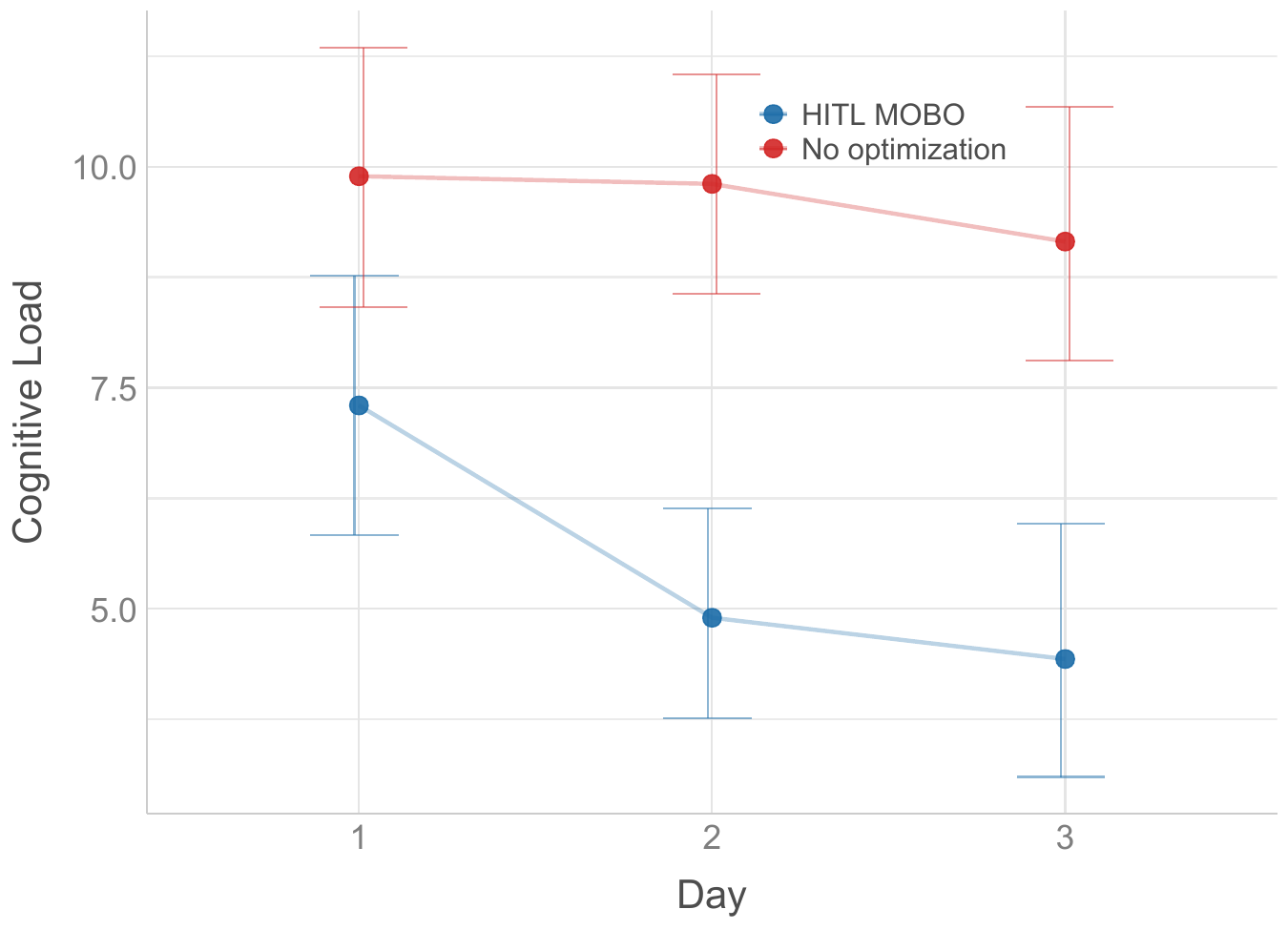}
        \caption{IE \longitudinalBo $\times$ \run on cognitive load.}
        \Description{The graph shows the interaction effect of the day, combined with multi-session Bayesian optimization, on mental workload. When Bayesian Optimization was used, the mental workload was significantly lower.}
        \label{fig:ie_long_run}
    \end{minipage}
    \hfill
    \begin{minipage}[t]{0.48\textwidth}
        \centering
        \includegraphics[width=0.8\linewidth]{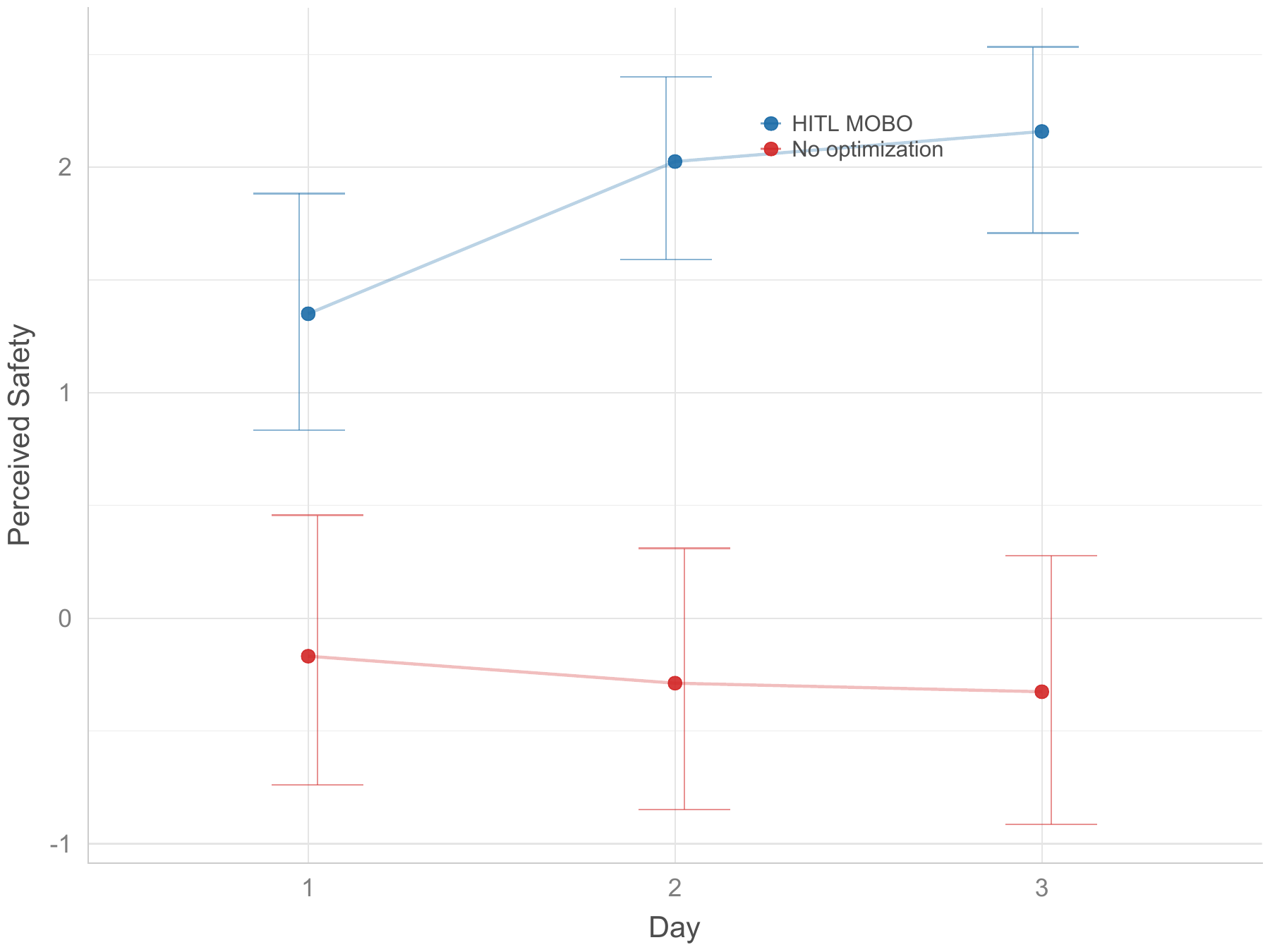}
        \caption{IE \longitudinalBo $\times$ \run on perceived safety.}
        \Description{The graph shows the interaction effect of the day, combined with multi-session Bayesian optimization, on perceived safety. When Bayesian Optimization was used, perceived safety was significantly higher.}
        \label{fig:ie_long_run_ps}
    \end{minipage}
\end{figure*}

The ART found a significant main effect of \longitudinalBo (\F{1}{65}{13.47}, \pminor{0.001}), of \run (\F{2}{130}{9.24}, \pminor{0.001}), and a significant interaction effect (IE) of \longitudinalBo $\times$ \run on cognitive load (\F{2}{130}{3.98}, \p{0.021}; see \autoref{fig:ie_long_run}).
%While cognitive load was almost always lower with HITL MOBO, with the expert designs, it was comparable to the conditions without (see \autoref{fig:ie_long_group_tlx}). 
\autoref{fig:ie_long_run} shows that with HITL MOBO, the cognitive load was always lower and also had a steeper decline over the course of the three days.
ART contrasts showed cognitive load decreased from day one to day two ($t(100)=2.50$, \padj{0.028}) across both groups and from day one to day three ($t(100)=3.76$, \padjminor{0.001}), with no further day two-to-three change.

\paragraph{Trust}

\begin{figure*}[ht!]
\centering
    \begin{subfigure}[b]{0.49\linewidth}
        \centering
        \includegraphics[width=0.8\linewidth]{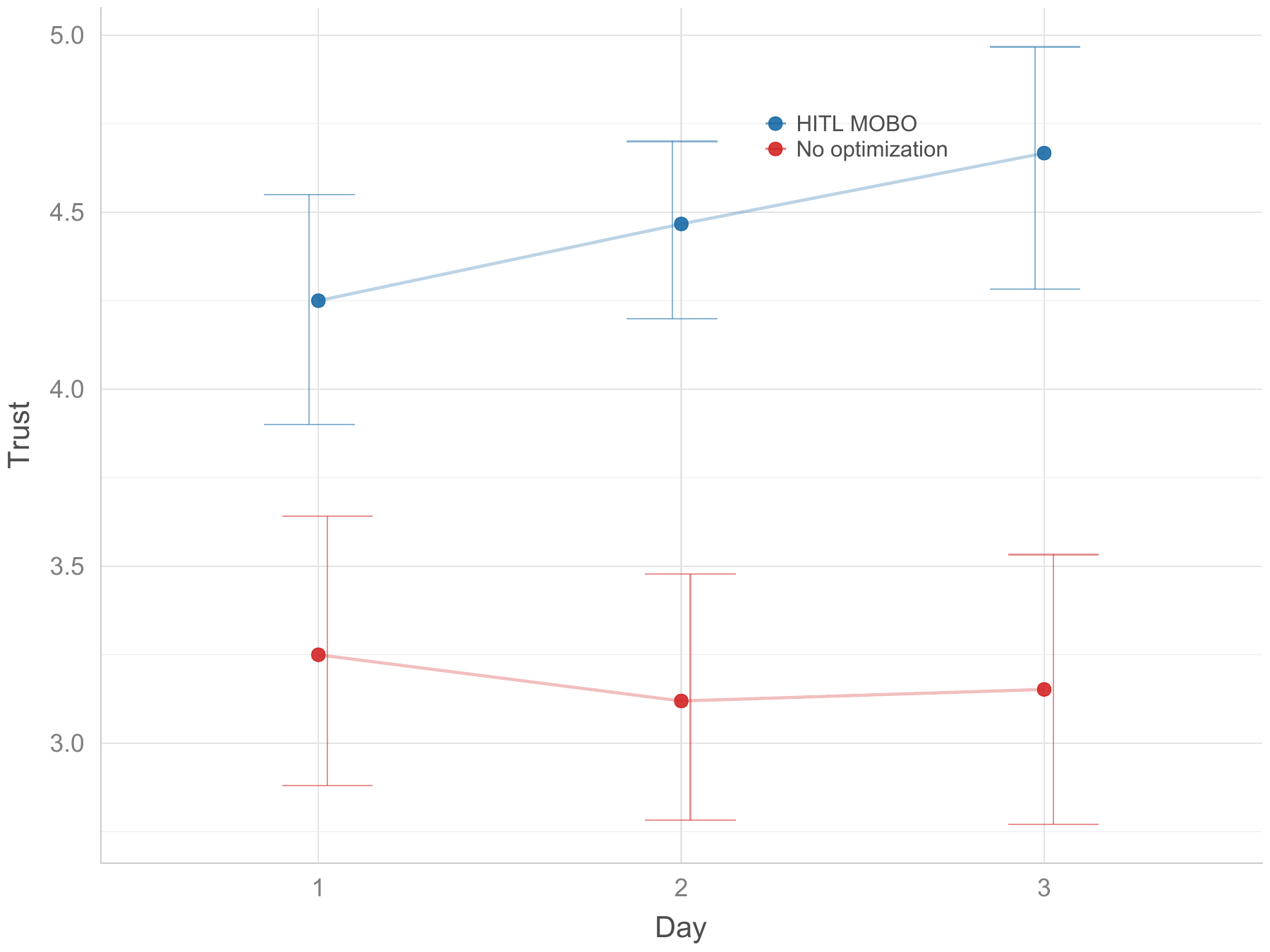}
        \caption{IE \longitudinalBo $\times$ \run on trust.}
        \label{fig:ie_long_run_trust}
        \Description{Interaction effect of day and multi-session Bayesian optimization on trust. With HITL MOBO, trust was higher and continued to rise across the three days.}
    \end{subfigure}
    \hfill
    \begin{subfigure}[b]{0.49\linewidth}
        \centering
        \includegraphics[width=0.8\linewidth]{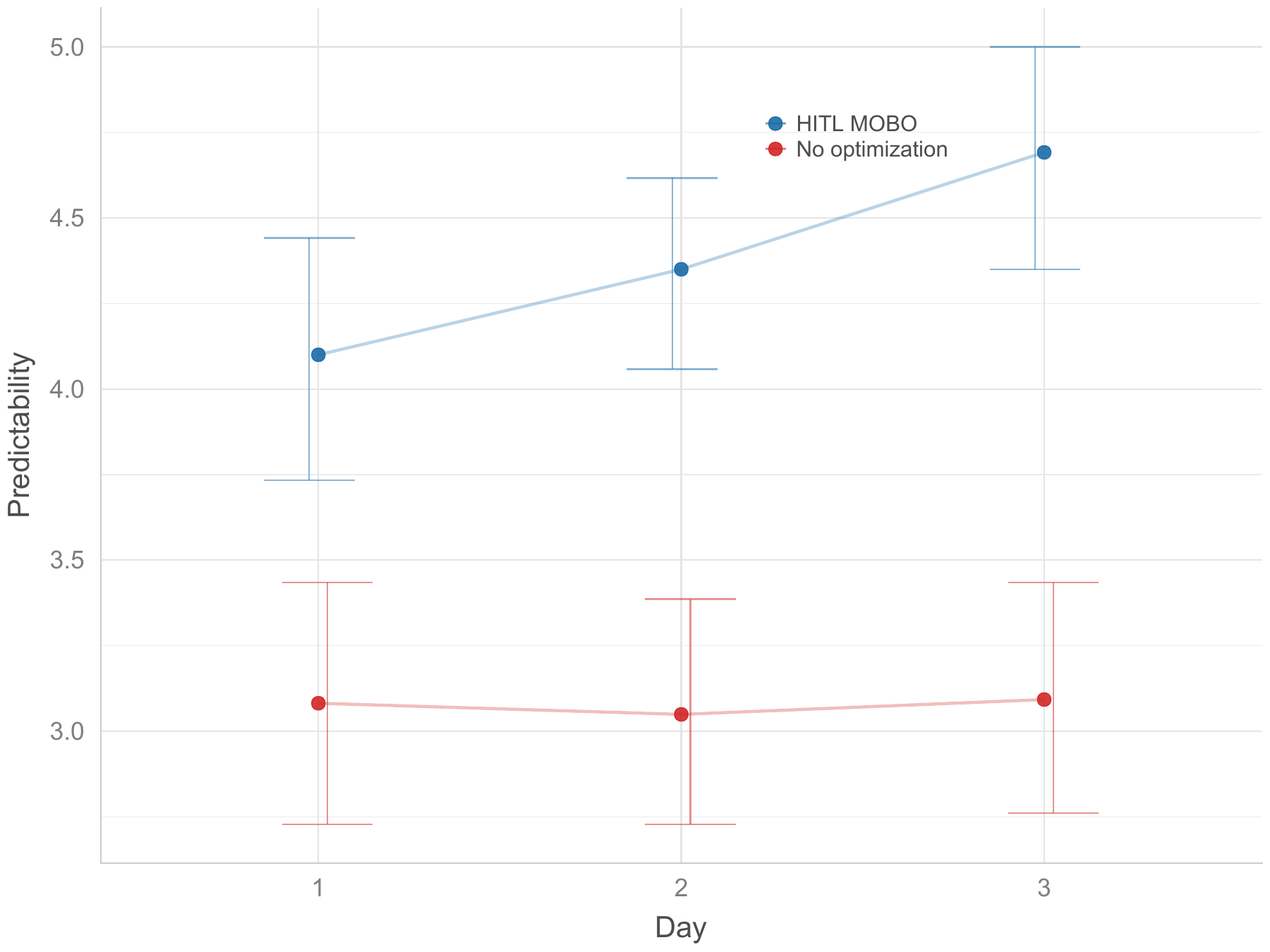}
        \caption{IE \longitudinalBo $\times$ \run on predictability.}
        \label{fig:ie_long_run_pred}
        \Description{Interaction effect of day and multi-session Bayesian optimization on predictability. With HITL MOBO, predictability was higher and continued to rise across the three days.}
    \end{subfigure}
    \caption{IEs on trust and predictability.}
    \Description{Two line graphs of trust and predictability across three days, split by optimization condition.}
\end{figure*}

The ART found a significant main effect of \GroupID (\F{5}{65}{3.41}, \p{0.009}) and of \longitudinalBo on trust (\F{1}{65}{23.42}, \pminor{0.001}). 
ART contrasts (Holm-corrected over C2--C6) found that C4-Cold-Start HITL MOBO was significantly higher in \trust than C2-Custom design by experts ($t(50)=3.67$, \padj{0.006}); no other condition pair differed significantly (see \autoref{fig:ie_long_run_trust}).

% old DUnn test
%A post-hoc test found that C4-Cold-Start HITL MOBO was significantly higher (\m{4.33}, \sd{0.92}) in terms of \trust compared to C2-Custom design by experts (\m{2.99}, \sd{1.38}; \padjminor{0.001}) and than C1-No Vis. (\m{3.42}, \sd{1.09}; \padj{0.003}). C5-Expert-Informed Warm-Start HITL MOBO was significantly higher (\m{4.09}, \sd{1.51}) C2-Custom design by experts (\m{2.99}, \sd{1.38}; \padj{0.003}). With \longitudinalBo, trust was significantly higher (\m{4.46}, \sd{0.88}) than without (\m{3.17}, \sd{1.29}). 

\paragraph{Predictability}

The ART found a significant main effect of \GroupID (\F{5}{65}{3.13}, \p{0.014}) and of \longitudinalBo on predictability (\F{1}{65}{24.26}, \pminor{0.001}). 

ART contrasts found that C4-Cold-Start HITL MOBO was significantly higher in \predictability than C2-Custom design by experts ($t(50)=3.09$, \padj{0.033}); no other condition pair differed significantly (see \autoref{fig:ie_long_run_pred}).

%A post-hoc test found that C4-Cold-Start HITL MOBO was significantly higher (\m{4.24}, \sd{0.93}) in terms of \predictability compared to C2-Custom design by experts (\m{3.11}, \sd{1.35}; \padjminor{0.001}), than C1-No Vis. (\m{3.26}, \sd{0.93}; \padjminor{0.001}), and than C3-Custom design by end-users (\m{3.47}, \sd{1.25}; \padj{0.021}).  With \longitudinalBo, predictability was significantly higher (\m{4.38}, \sd{0.95}) than without (\m{3.07}, \sd{1.16}).

\paragraph{Perceived Safety}

The ART found a significant main effect of \GroupID (\F{5}{65}{2.95}, \p{0.018}) and of \longitudinalBo on \perceivedSafetyScore (\F{1}{65}{20.88}, \pminor{0.001}).  The ART also found a significant IE of \longitudinalBo $\times$ \run on \perceivedSafetyScore (\F{2}{130}{3.19}, \p{0.044}; see \autoref{fig:ie_long_run_ps}). 

ART contrasts found that C4-Cold-Start HITL MOBO was significantly higher in \perceivedSafetyScore than both C2-Custom design by experts ($t(50)=3.07$, \padj{0.035}) and C3-Custom design by end-users ($t(50)=2.94$, \padj{0.045}); no other condition pair differed significantly.

%A post-hoc test found that C4-Cold-Start HITL MOBO was significantly higher (\m{1.55}, \sd{1.61}) in terms of \perceivedSafetyScore compared to C2-Custom design by experts (\m{-0.15}, \sd{1.83}; \padj{0.002}), than C1-No Vis. (\m{0.32}, \sd{2.03}; \padj{0.046})
%A post-hoc test found that C5-Expert-Informed Warm-Start HITL MOBO was significantly higher (\m{1.44}, \sd{2.07}) in terms of \perceivedSafetyScore compared to C2-Custom design by experts (\m{-0.15}, \sd{1.83}; \padj{0.013}). A post-hoc test found that C4-Cold-start HITL MOBO was significantly higher (\m{1.55}, \sd{1.61}) in terms of \perceivedSafetyScore compared to C3-Custom design by end-users (\m{-0.24}, \sd{2.29}; \padj{0.002}). A post-hoc test found that C5-Expert-Informed Warm-Start HITL MOBO was significantly higher (\m{1.44}, \sd{2.07}) in terms of \perceivedSafetyScore compared to C3-Custom design by end-users (\m{-0.24}, \sd{2.29}; \padj{0.013}). 
%and C3-Custom design by end-users (\m{-0.22}, \sd{2.29}; \padj{0.002}). A post-hoc test found that C5-Expert-Informed Warm-Start HITL MOBO was significantly higher (\m{1.34}, \sd{2.18}) in terms of \perceivedSafetyScore compared to C3-Custom design by end-users (\m{-0.22}, \sd{2.29}; \padj{0.027}). 

%With \longitudinalBo, perceived safety was significantly higher (\m{1.84}, \sd{1.32}) than without (\m{-0.26}, \sd{2.09}). 
\autoref{fig:ie_long_run_ps} shows that, without optimization, perceived safety was almost equal across the three days; with the HITL MOBO, the difference between Days 1 and 2 is the largest.

\paragraph{Aesthetics}

\begin{figure*}[ht!]
\centering
         \begin{subfigure}[b]{0.49\linewidth}
    \includegraphics[width=0.8\linewidth]{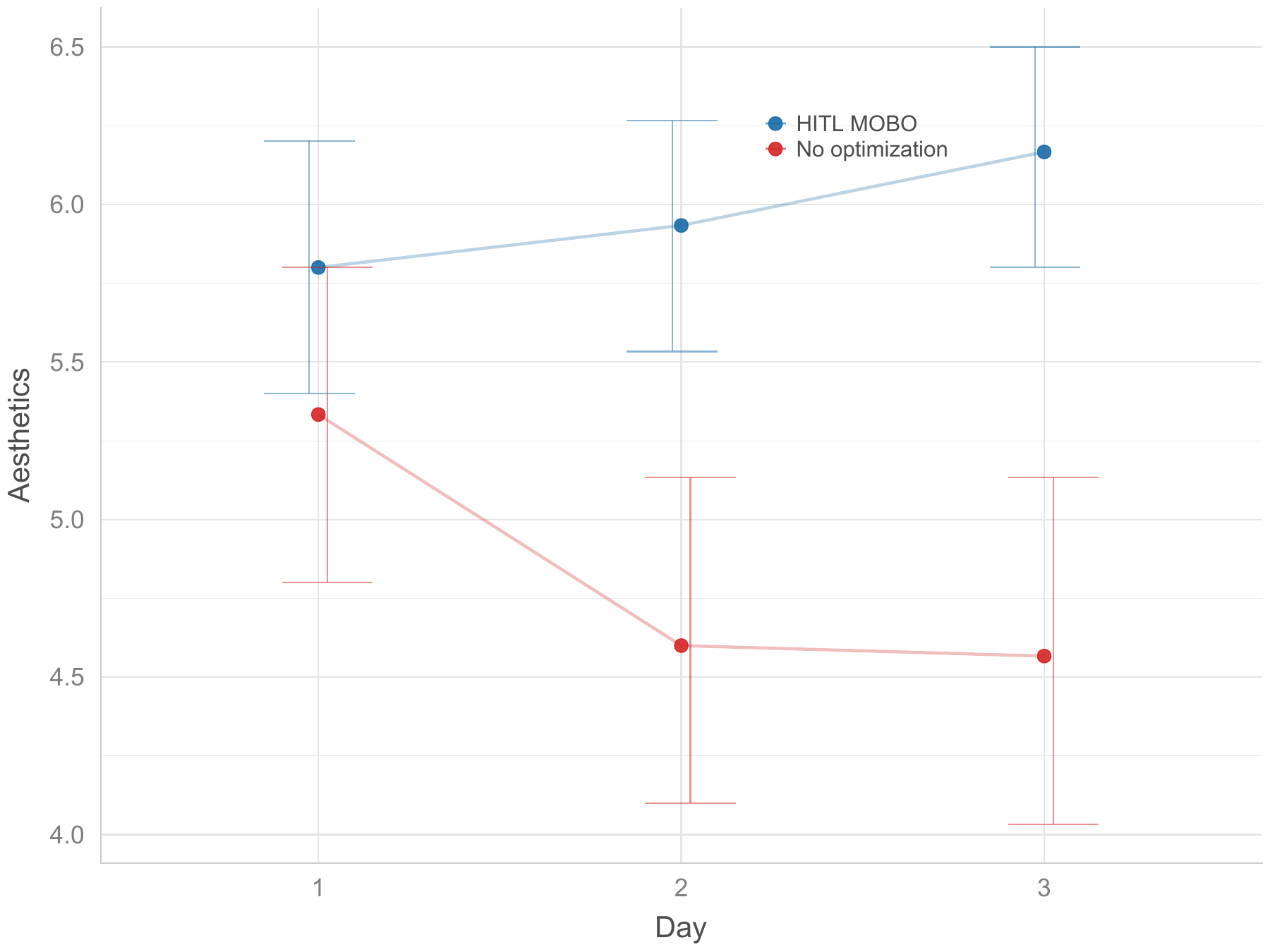}
   \caption{IE \run $\times$ \longitudinalBo on aesthetics.}
   \label{fig:ie_long_run_aesthetics}
    \Description{The graph shows the interaction effect of the day, combined with multi-session Bayesian optimization, on aesthetics. When Bayesian Optimization was used, the aesthetics were significantly higher.}
      \end{subfigure}
         \begin{subfigure}[b]{0.49\linewidth}
    \includegraphics[width=0.8\linewidth]{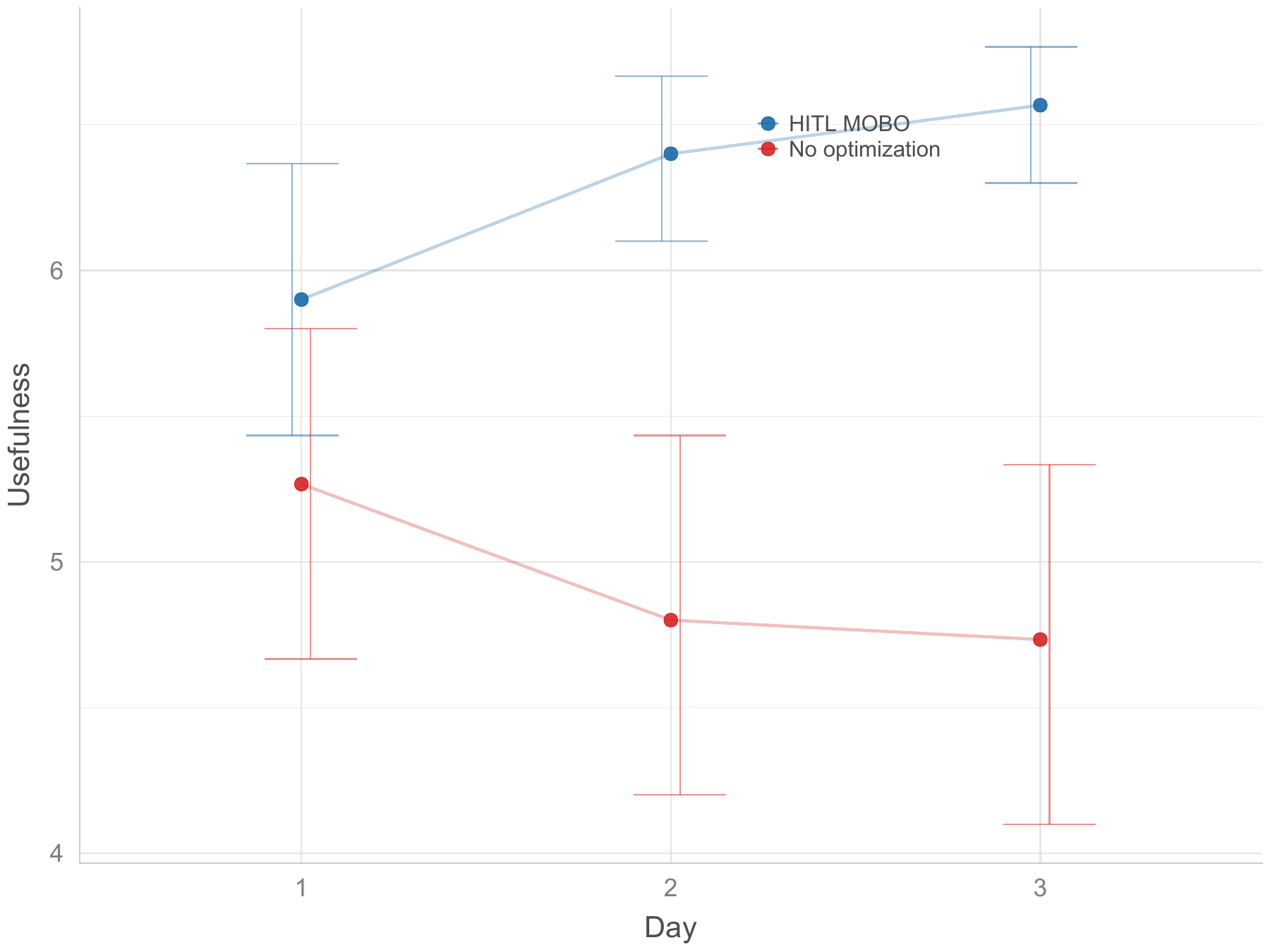}
   \caption{IE \run $\times$ \longitudinalBo on usefulness.}
   \label{fig:ie_long_run_usefulness}
    \Description{The graph shows the interaction effect of the day, combined with multi-session Bayesian optimization, on usefulness. When Bayesian Optimization was used, the usefulness was significantly higher.}
      \end{subfigure}
      \caption{IEs on aesthetics and usefulness.}
      \Description{The graph shows the interaction effects on usefulness and aesthetics. When Bayesian Optimization was used, both improved over time.}
\end{figure*}

The ART found a significant main effect of \GroupID (\F{4}{50}{2.77}, \p{0.037}) and of \longitudinalBo on aesthetics (\F{1}{50}{12.92}, \pminor{0.001}). The ART also found a significant IE of \longitudinalBo $\times$ \run on aesthetics (\F{2}{100}{7.66}, \pminor{0.001}; see \autoref{fig:ie_long_run_aesthetics}). 

Although the omnibus effect of \GroupID was significant, no pairwise condition contrast survived Holm correction (all adjusted $p>.05$); the day-level pattern is shown in \autoref{fig:ie_long_run_aesthetics}.

With \longitudinalBo, aesthetics was higher (\m{5.97}, \sd{1.05}) than without (\m{4.83}, \sd{1.53}).

%A post-hoc test found that C4-Cold-Start HITL MOBO was significantly higher (\m{5.82}, \sd{1.32}) in terms of \Aesthetics compared to C2-Custom design by experts (\m{4.84}, \sd{1.36}; \padj{0.006}) and than C3-Custom design by end-users (\m{4.95}, \sd{1.43}; \padj{0.027})
%A post-hoc test found that C6-User-Informed Warm-Start HITL MOBO was significantly higher (\m{5.88}, \sd{0.99}) in terms of \Aesthetics compared to C2-Custom design by experts (\m{4.84}, \sd{1.36}; \padj{0.035}). 
%A post-hoc test found that C5-Expert-Informed Warm-Start HITL MOBO was significantly higher (\m{5.85}, \sd{1.59}) in terms of \Aesthetics compared to C2-Custom design by experts (\m{4.84}, \sd{1.36}; \padj{0.007}) and than C3-Custom design by end-users (\m{4.95}, \sd{1.43}; \padj{0.030}).

\autoref{fig:ie_long_run_aesthetics} shows that, while on day one aesthetics was only a bit higher for the conditions that would have HITL MOBO, over the three days aesthetics was rated even better, whereas it declined severely without HITL MOBO.

\paragraph{Usefulness}

The ART found a significant main effect of \longitudinalBo on Usefulness (\F{1}{50}{14.36}, \pminor{0.001}) and a significant IE of \longitudinalBo $\times$ \run on Usefulness (\F{2}{100}{6.67}, \p{0.002}; see \autoref{fig:ie_long_run_usefulness}). On day 1, usefulness was only a bit higher for the conditions that would have HITL MOBO. Over the three days, usefulness was rated even better, while it declined rather severely without HITL MOBO.

With \longitudinalBo, usefulness was higher (\m{6.29}, \sd{1.06}) than without (\m{4.93}, \sd{1.69}).

\paragraph{Satisfying}

\begin{figure*}[ht!]
\centering
    \begin{minipage}[t]{0.48\textwidth}
        \centering
        \includegraphics[width=.8\linewidth]{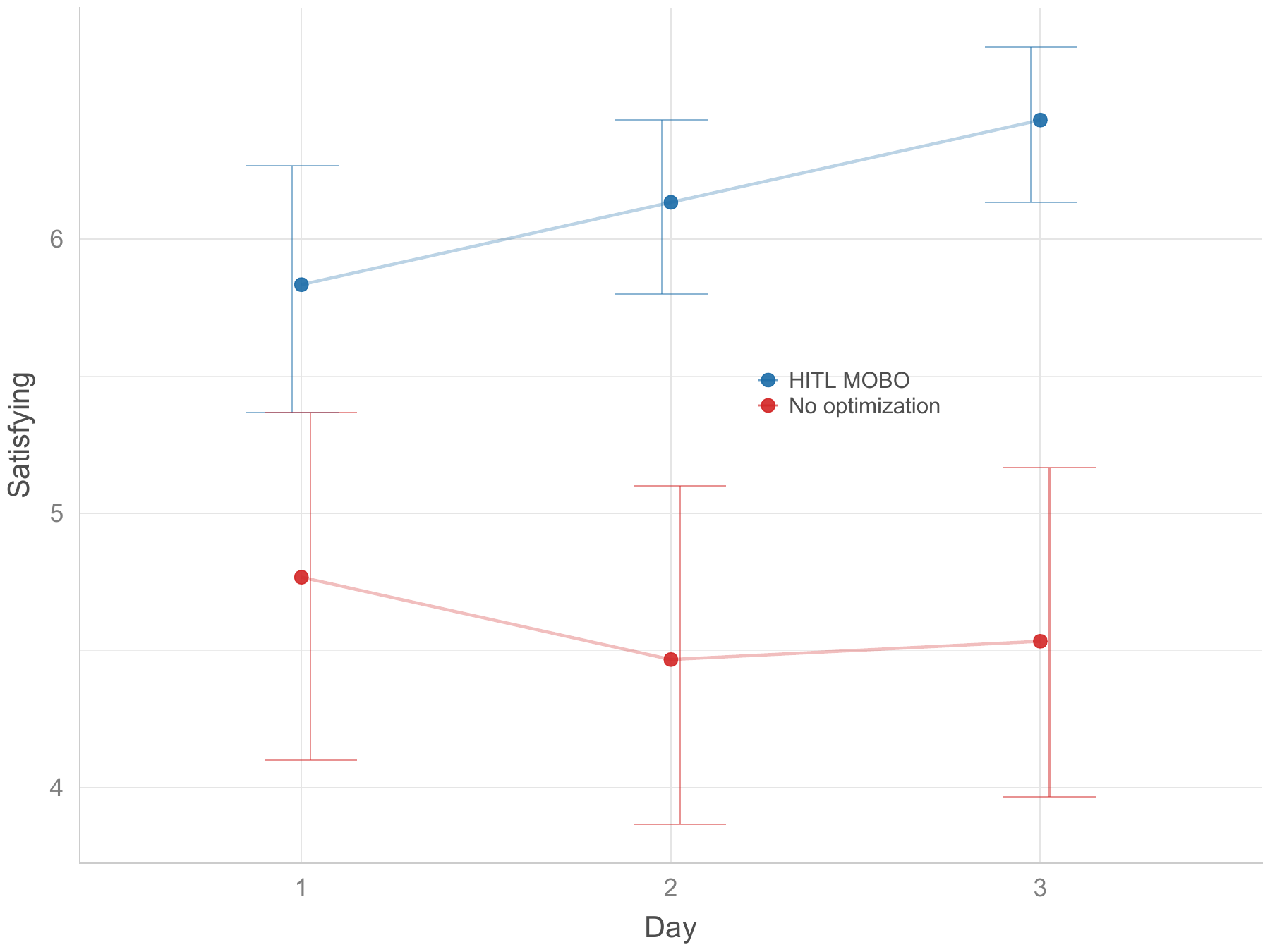}
        \caption{IE \longitudinalBo $\times$ \run on satisfying.}
        \Description{The graph shows the interaction effect of the day combined with multi-session Bayesian optimization on satisfying. When Bayesian Optimization was used, satisfying was significantly higher.}
        \label{fig:ie_long_run_sat}
    \end{minipage}
    \hfill
    \begin{minipage}[t]{0.48\textwidth}
        \centering
        \includegraphics[width=.8\linewidth]{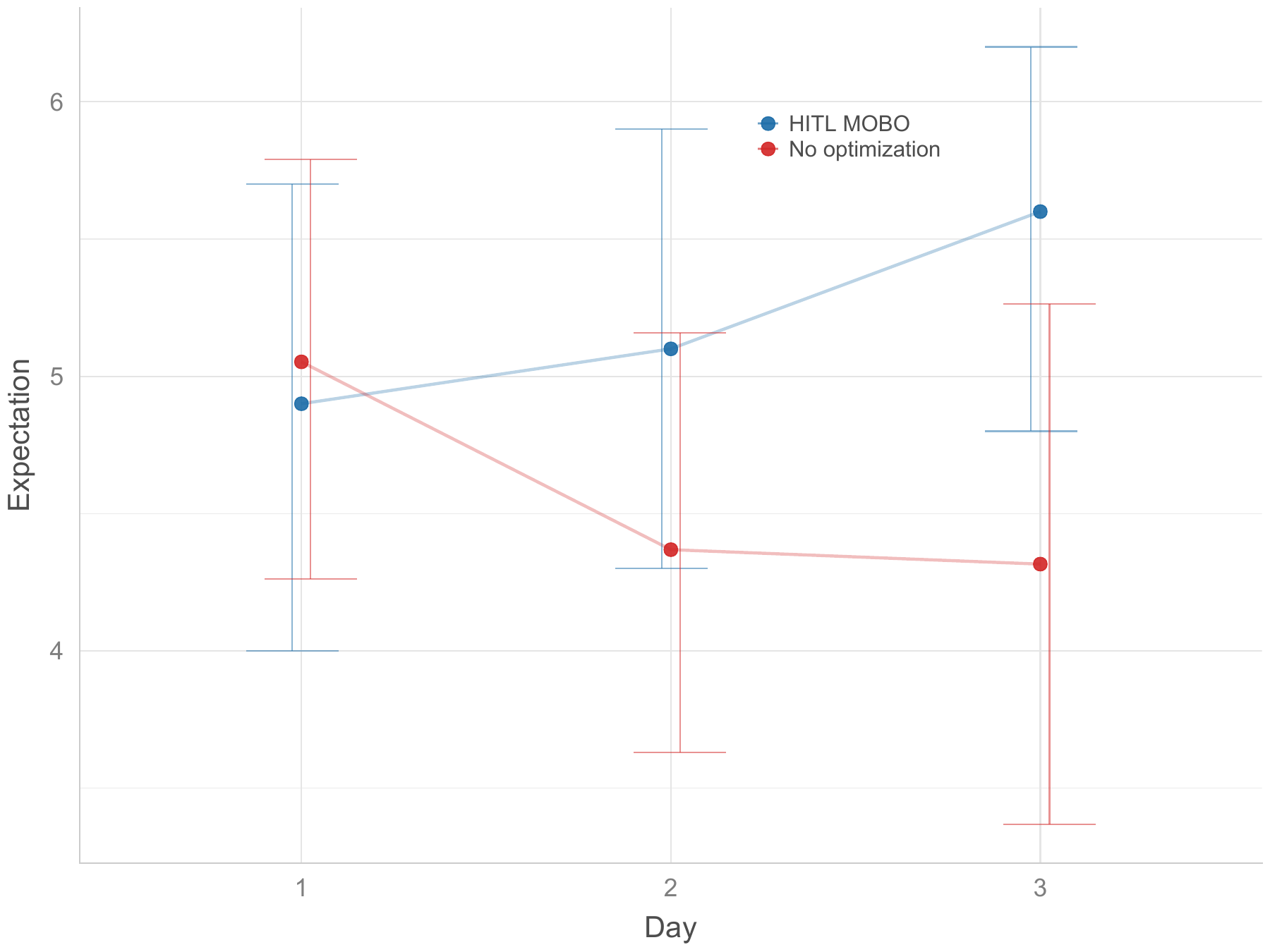}
        \caption{IE \longitudinalBo $\times$ \run on expectation conformity.}
        \Description{The graph shows the interaction effect of the day combined with multi-session Bayesian optimization on expectation conformity. When Bayesian Optimization was used, the expectation conformity was significantly higher.}
        \label{fig:expectation_longitudinal_interaction}
    \end{minipage}
\end{figure*}

The ART found a significant main effect of \GroupID (\F{4}{50}{4.51}, \p{0.003}) and of \longitudinalBo on satisfying (\F{1}{50}{18.41}, \pminor{0.001}). The ART also found a significant IE of \longitudinalBo $\times$ \run on satisfying (\F{2}{100}{3.76}, \p{0.027}; see \autoref{fig:ie_long_run_sat}). 

ART contrasts found that C4-Cold-Start HITL MOBO was significantly higher in \AcceptanceTwo than both C2-Custom design by experts ($t(50)=3.38$, \padj{0.014}) and C3-Custom design by end-users ($t(50)=3.39$, \padj{0.014}); no other condition pair differed significantly.

%A post-hoc test found that C4-Cold-Start HITL MOBO was significantly higher (\m{6.16}, \sd{1.30}) in terms of \AcceptanceTwo compared to C2-Custom design by experts (\m{4.80}, \sd{1.56}; \padjminor{0.001}) and than C3-Custom design by end-users (\m{4.62}, \sd{1.63}; \padjminor{0.001}). A post-hoc test found that C5-Expert-Informed Warm-Start HITL MOBO was significantly higher (\m{5.67}, \sd{1.80}) in terms of \AcceptanceTwo compared to C3-Custom design by end-users (\m{4.62}, \sd{1.63}; \padj{0.023}). 

With \longitudinalBo, satisfying was higher (\m{6.13}, \sd{1.05}) than without (\m{4.59}, \sd{1.71}).

\paragraph{User Expectation}
The ART found a significant interaction effect of \longitudinalBo $\times$ \run on expectation conformity (\F{2}{40}{3.60}, \p{0.037}; see \autoref{fig:expectation_longitudinal_interaction}). Over the three days, the design was much closer to expectations with \longitudinalBo.  

\paragraph{Satisfaction}
The ART found a significant main effect of \run on satisfaction (\F{2}{40}{7.21}, \p{0.002}). 
Pairwise day contrasts were not estimable in the full model owing to the sparse design-experience cells.

\paragraph{Confidence}
The ART found no significant effects on confidence in the design parameters. 

\paragraph{Agency}

\begin{figure}[ht!]
\centering
    \includegraphics[width=0.4\linewidth]{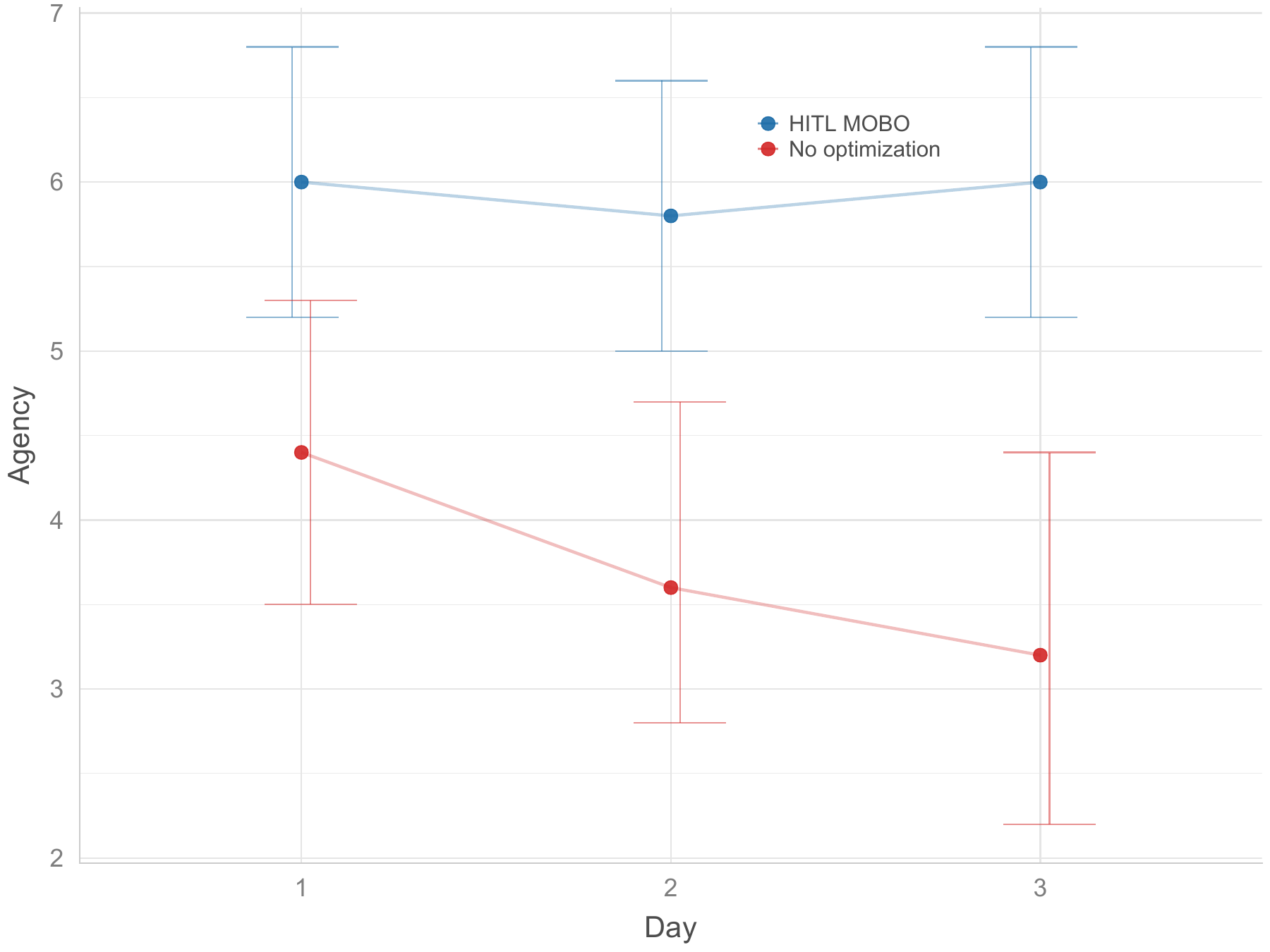}
   \caption{IE \longitudinalBo $\times$ \run on agency.}
       \Description{Agency by day and optimization.}
   \label{fig:agency_longitudinal_interaction_bo_run}
\end{figure}

Because the design-experience items were collected only for the design-process conditions and yielded sparse, unbalanced \GroupID cells, we omit \GroupID from these models and test \longitudinalBo and \run (and their interaction). The ART found a significant main effect of \longitudinalBo (\F{1}{13}{7.14}, \p{0.019}) and of \run on agency (\F{2}{26}{4.90}, \p{0.016}); the \longitudinalBo $\times$ \run interaction was not significant (\F{2}{26}{2.77}, \p{0.081}). \autoref{fig:agency_longitudinal_interaction_bo_run} shows that agency was higher with \longitudinalBo and declined over the three days, more so without optimization.

\paragraph{Ownership}
As for agency, we test \longitudinalBo and \run without \GroupID. The ART found a significant main effect of \longitudinalBo (\F{1}{13}{6.98}, \p{0.020}) and of \run on ownership (\F{2}{26}{6.64}, \p{0.005}). With \longitudinalBo, ownership was significantly higher (\m{6.00}, \sd{0.93}) than without (\m{3.80}, \sd{1.75}). ART contrasts showed ownership was higher on day one than on day two ($t(26)=3.42$, \padj{0.006}) and day three ($t(26)=3.04$, \padj{0.011}), with no day two-to-three difference.

\subsection{Optimization Dynamics Across Sessions}\label{sec:dynamics}

\begin{figure*}[ht!]
\centering
    \begin{subfigure}[b]{0.32\linewidth}
        \centering
        \includegraphics[width=\linewidth]{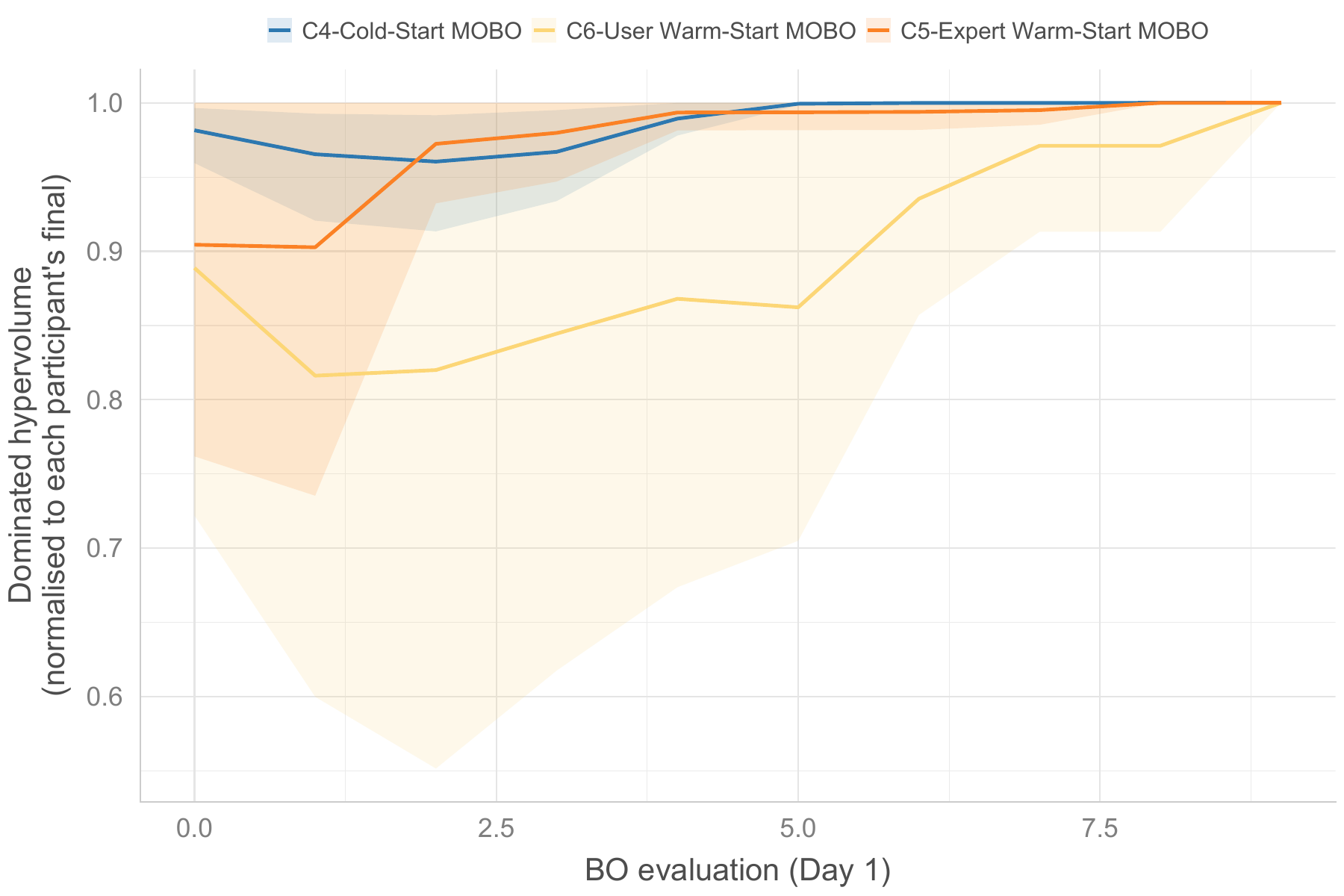}
        \caption{Day-one convergence: within-participant normalized hypervolume per BO evaluation.}
        \label{fig:dyn_convergence}
        \Description{Hypervolume convergence curves per MOBO condition on day one. Warm starts begin near their converged value; the cold start improves over more evaluations.}
    \end{subfigure}
    \hfill
    \begin{subfigure}[b]{0.32\linewidth}
        \centering
        \includegraphics[width=\linewidth]{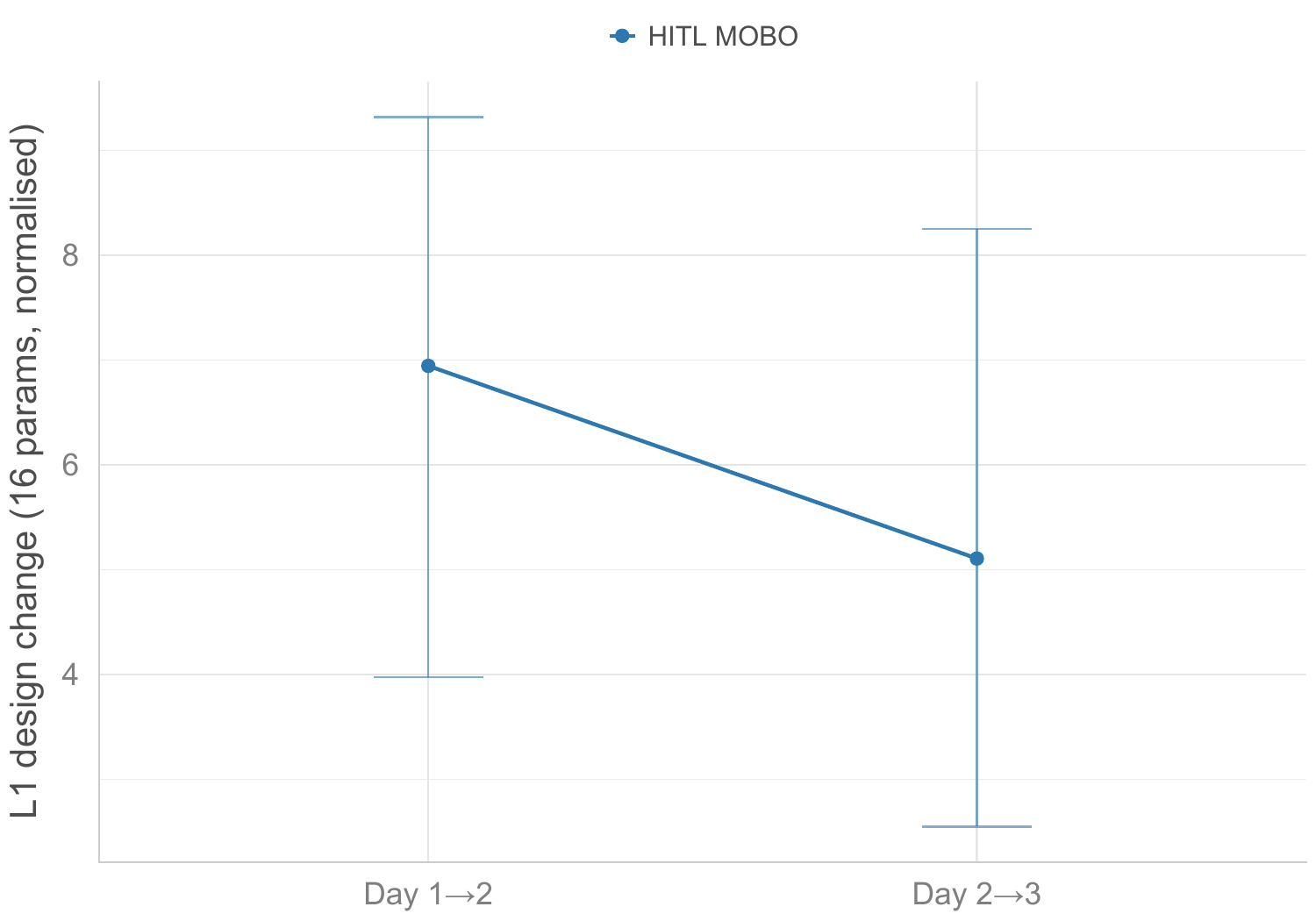}
        \caption{Day-to-day design drift: L1 change over the 16 parameters by arm.}
        \label{fig:dyn_drift}
        \Description{L1 parameter change between consecutive days. The frozen no-optimization arm is near zero; the continued-MOBO arm is larger.}
    \end{subfigure}
    \hfill
    \begin{subfigure}[b]{0.32\linewidth}
        \centering
        \includegraphics[width=\linewidth]{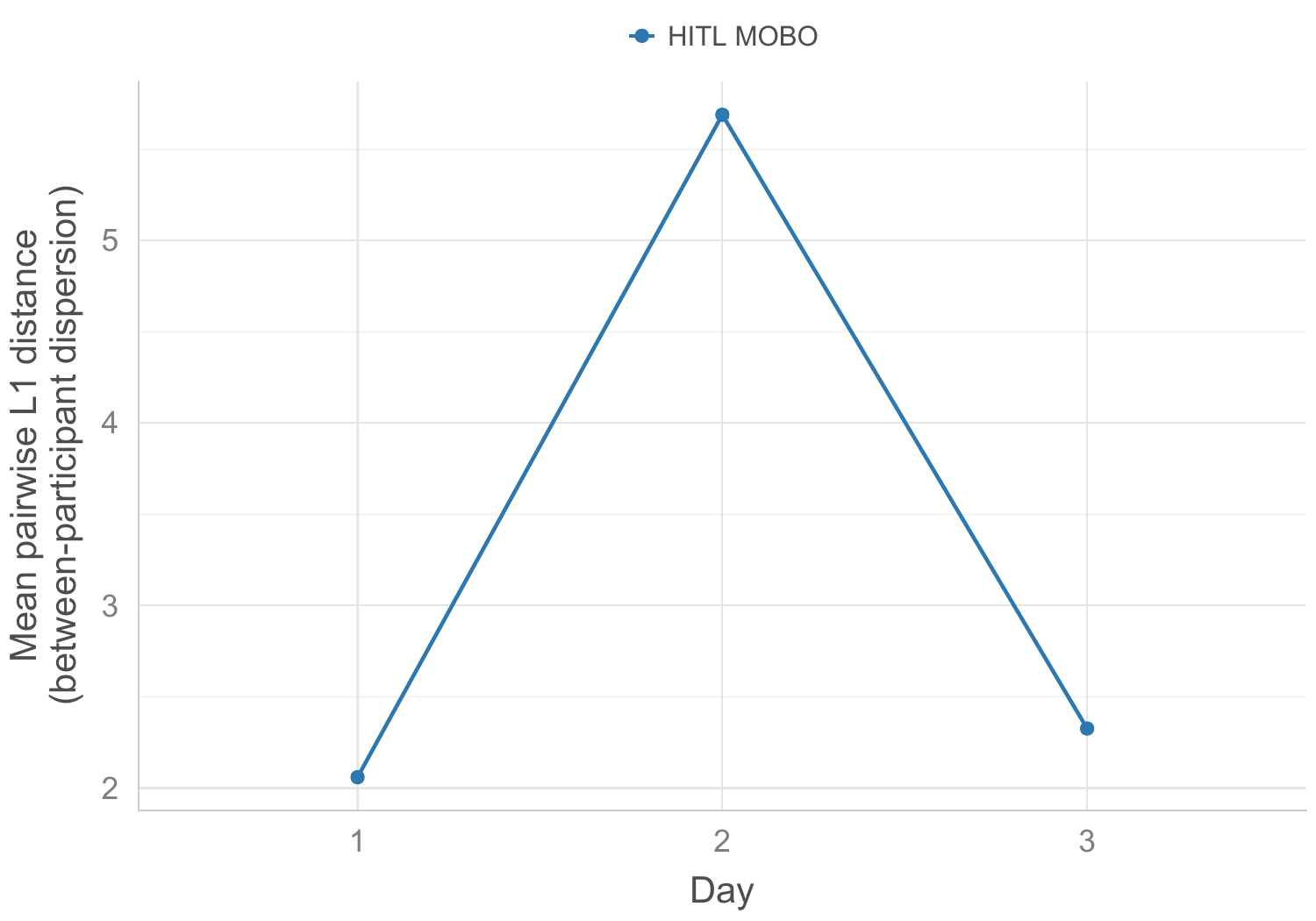}
        \caption{Personalization: between-participant dispersion of final designs per day.}
        \label{fig:dyn_dispersion}
        \Description{Between-participant dispersion of designs across days, showing the optimizer converges to personalized rather than uniform designs.}
    \end{subfigure}
    \caption{Optimization dynamics across the multi-session deployment (RQ3): day-one convergence~(\subref{fig:dyn_convergence}), day-to-day design drift~(\subref{fig:dyn_drift}), and between-user personalization~(\subref{fig:dyn_dispersion}). We treat the parameter-space panels as preliminary because per-day logs are incompletely recorded for part of the sample (see Limitations).}
    \label{fig:dynamics}
    \Description{Three panels summarizing convergence, design drift, and personalization dispersion across sessions.}
\end{figure*}

Beyond the per-objective outcomes, we characterize \emph{how} the optimization behaves across the multi-session deployment (RQ3).

\paragraph{Day-one convergence.}
Using the hypervolume logged per BO evaluation, the warm-start conditions started close to their converged value and stabilized in few evaluations (e.g., the expert-informed C5 reached $\approx$95\% of its day-one hypervolume gain within $\approx$7 evaluations; see \autoref{fig:dyn_convergence}), whereas the cold-start C4 began with more head-room and improved over more evaluations. This is consistent with warm-starting narrowing the search and with the absence of a final-design advantage for the warm-starts: the additional starting information shortens the path rather than reaching a better optimum.

\begin{table}[ht]
\caption{Within the continued-MOBO arm: per-day means and consecutive-day deltas,
classifying each objective as front-loaded (novelty-like, almost all the change happens in the first step) or sustained.}
\label{tab:novelty}\footnotesize
\begin{tabular}{@{}lcccccl@{}}
\toprule
\textbf{Objective} & \textbf{D1} & \textbf{D2} & \textbf{D3} & \textbf{$\Delta_{1\to2}$} & \textbf{$\Delta_{2\to3}$} & \textbf{Pattern} \\
\midrule
Cognitive load   & 7.3 & 4.9 & 4.4 & $-2.4$ & $-0.5$ & front-loaded \\
Perceived safety & 1.4 & 2.0 & 2.2 & $0.7$  & $0.1$  & front-loaded \\
Usefulness       & 5.9 & 6.4 & 6.6 & $0.5$  & $0.2$  & front-loaded \\
Trust            & 4.3 & 4.5 & 4.7 & $0.2$  & $0.2$  & sustained \\
Predictability   & 4.1 & 4.4 & 4.7 & $0.3$  & $0.3$  & sustained \\
Aesthetics       & 5.8 & 5.9 & 6.2 & $0.1$  & $0.2$  & sustained \\
Satisfying       & 5.8 & 6.1 & 6.4 & $0.3$  & $0.3$  & sustained \\
\bottomrule
\end{tabular}
\end{table}

\paragraph{Novelty vs.\ sustained gains.}
Within the continued-MOBO arm, consecutive-day ART contrasts reveal two patterns (see \autoref{tab:novelty}). Cognitive load decreased mainly from day one to day two ($t(58)=4.41$, \pminor{0.001}) and perceived safety increased mainly from day one to day two ($t(58)=3.33$, \padj{0.003}), each with no further day two-to-three change---a front-loaded, novelty-like pattern. In contrast, trust, predictability, aesthetics, and satisfying continued to rise to day three. Thus, part of the benefit reflects early repeated-exposure effects, while several constructs improve in a genuinely sustained manner.

\paragraph{Design drift and personalization (preliminary).}
Per-participant day-to-day change in the 16 design parameters was, as expected, near zero for the frozen (\emph{no optimization}) arm and larger for the continued-MOBO arm, confirming the manipulation (see \autoref{fig:dyn_drift}). Between-participant dispersion of the final designs remained substantial across days rather than collapsing to a single solution, indicating that the optimizer converged to \emph{personalized} rather than uniform designs (see \autoref{fig:dyn_dispersion}). We report these design-space analyses as preliminary because the per-day parameter logs are incompletely recorded for part of the sample (see Limitations).

%%%%%%%%%%%%%%%%%%%%%%%%%%%%%%%%%%%%%%
\subsection{Final Parameter Set} 

Appendix \autoref{fig:final_params} shows the parameters over all three days with regard to the different strategies. While the overall trend remains the same (e.g., the HUD value remains the same), some differences are notable. For example, the HUD's alpha value changed over time. First, it increased, then it decreased. While this does not constitute an in-depth analysis of changes over the multi-session study, it shows that the MOBO process worked as intended.

\section{Discussion}
This paper used the implemented in-vehicle functionality visualization conditions proposed by Jansen \& Colley et al. \cite{jansen2025opticarvis} (C2-C6, see Section~\ref{study-optimization-strategies}) and evaluated their impact over three days. Participants were split into two additional categories: with \longitudinalBo and with \textit{no optimization}.

\subsection{Multi-Session Exposure to Bayesian Optimization Improves Designs}

While previous work found that optimization strategies, in particular, the C4 Cold-Start HITL MOBO~\cite{jansen2025opticarvis} led to the best results, our work shows that these improvements are only consistent \textbf{when HITL MOBO remains activated}. This was the case for all conditions with MOBO (C4 - C6) and across various objectives (aesthetics, usefulness, cognitive load, perceived safety, trust, predictability, expectation conformity, and agency).
After correction, condition differences were driven mainly by C4 (cold-start) outperforming the static custom designs (C2/C3); the warm-start variants (C5/C6) were not reliably distinguishable from the custom designs, and aesthetics/usefulness showed no surviving pairwise condition differences.
 Therefore, our data show that MOBO is an appropriate method for adapting in-vehicle functionality visualizations over time but more work is needed to understand the role of warm-start variants.

\subsection{The Need for Implicit Design Optimizations of In-Vehicle Visualizations}

Our MOBO application currently relies on an explicit optimization loop per day that frequently queries users about their subjective perception of the AV. While we found that querying once daily can improve users' subjective perceptions, this method has drawbacks, including user fatigue and potential inaccuracies in feedback (see \citet{chan2022bo}).

%Taking a cue from \citet{koyama2022boassistant}, we are considering a shift towards an implicit feedback mechanism. 
Implicit feedback mechanisms could resolve these challenges (see \citet{koyama2022boassistant}). This involves gathering additional passive feedback, such as user interactions, physiological markers (e.g., heart rate), and psychological states. Such an approach aligns with the non-intrusive methods discussed in research by \citet{10.1145/3546726} and \citet{colley2024autotherm}. However, this has challenges, such as the diversity of participants and inaccuracies in situations. The presented methods also allow for incorporating prior (expert) knowledge via the \textit{Warm-Start} method (see C5 and C6).

\subsection{The Need for Customizability Despite Computational Optimization}

Even when satisfied with the optimized result, users in prior work still asked for explicit control over the design (see \citet{jansen2025opticarvis}).
In our use case, this should be easily incorporable. For example, a traditional settings menu could be provided additionally, upon which the optimizer can adapt the design parameters.

\subsection{Data-Driven Adaptability in Automated Vehicles: From Initial Use to Longer-Term Personalization}

% Insights into the user enable continuous adaptation and value generation for the user. No explicit design or over-the-air update is necessary; all can be done on the AV itself with current technology. 
% Self-reliant updates over the entire life-cycle are possible. 
% Long-term changes are no problem as rides are often longer. Also, car users identify with them and are happy to work with and on them.

% Life-long learning about the user could become a distinction for OEMs. With the ever-decreasing differentiation of physical appearance and the need for performance, personalization becomes more important. This is driven by data and its enabled personalization.

Incorporating MOBO into the design of AV functionality visualizations offers intriguing ways to continuously adapt to user needs. This self-sufficiency obviates the need for explicit design interventions or over-the-air updates, both of which can be implemented on the AV's existing computational infrastructure. This is particularly beneficial throughout the vehicle's life cycle, enabling it to adapt over the longer term. Given that car journeys often involve long periods, this provides a substantial window for optimizing functionality visualizations.
This aligns with frameworks such as LASR~\cite{rittger2022adaptive}, which emphasize adaptivity as a key factor in enhancing user experience throughout the vehicle’s lifecycle.

\subsection{Limitations and Future Work}
All dependent variables are subjective. We did not assess objective outcomes such as actual safety, situation awareness, or take-over performance; these are important complements for future work, particularly because the optimized design changes could affect monitoring behaviour.

Several constructs (cognitive load, aesthetics, usefulness, satisfying, and the design-experience items) were measured with single items and lack internal-reliability estimates. The optimizer searched a bounded set of 16 parameters at fixed positions; findings may not generalize to richer design spaces or to position/colour parameters held constant here.

The design-experience measures (expectation, satisfaction, confidence, agency, ownership) were collected only for the design-process conditions and analyzed on a small complete-case subsample ($N\approx15$); these results are exploratory.

Our three consecutive sessions capture \emph{early} multi-session use, not long-term (weeks/months) personalization. While three days of testing already represent an important advance over single-day studies, longer-term studies (e.g., over weeks or months of vehicle use) would provide further insights. As several gains were front-loaded (Section~\ref{sec:dynamics}), part of the benefit likely reflects novelty and repeated exposure; we therefore make no claims about long-term adaptation. Moreover, continuously changing an in-vehicle UI via over-the-air updates carries its own risks once a driver has habituated to a layout, and a more consistent ``one-size-fits-all'' design may be required to satisfy legislative or safety constraints. Balancing per-user optimization against such standardization is an open question.

The remote virtual study offered controlled but limited external validity. However, the complex and subjective nature of driving justified this initial approach for HITL MOBO. Future research should focus on real-world tests using technologies like XR-OOM~\cite{xroom}, PassengXR~\cite{PassengXR}, SwiVR Car-Seat~\cite{colley2021swivr}, or VAMPIRE~\cite{hock2022vampire}.
Visual clutter of visualizations remains a known challenge~\cite{colley2020effect, colley2022scene, colley2021effects}, which also occurred in this study when several visualizations were active simultaneously. 

Finally, the per-day design-parameter logs were incompletely recorded for part of the sample, and the day index used for the parameter trajectory is partly confounded with the optimization arm. We therefore treat the design-space analyses (Section~\ref{sec:dynamics}) and \autoref{fig:final_params} as descriptive/preliminary rather than as a clean within-subject trajectory.

\section{Conclusion}

In this multi-session study, we deployed MOBO to the complex design landscape of in-vehicle HMIs for AVs. The study involved 74 participants over three days and yielded compelling results. Our multi-session HITL MOBO approach was notably effective in adapting AV functionality visualizations regarding situation detection, situation prediction, and trajectory planning to users' preferences and needs. The optimization process had a pronounced impact on enhancing user trust, acceptance, perceived safety, and comprehension while simultaneously reducing cognitive load. These outcomes echo the findings in HCI literature that emphasize the importance of intuitive and adaptive visualizations in complex systems.

\section*{Open Science}
The Unity scenario will be available upon request. This will include installation instructions and information on required 3rd-party Unity assets.

\begin{acks}
We thank all study participants.
\end{acks}

\bibliographystyle{ACM-Reference-Format}
\bibliography{sample-base}

\appendix

\section{Participant Instructions} \label{appendix:participant-instructions}

\begin{figure*}[ht!]
\centering
    \includegraphics[width=\linewidth]{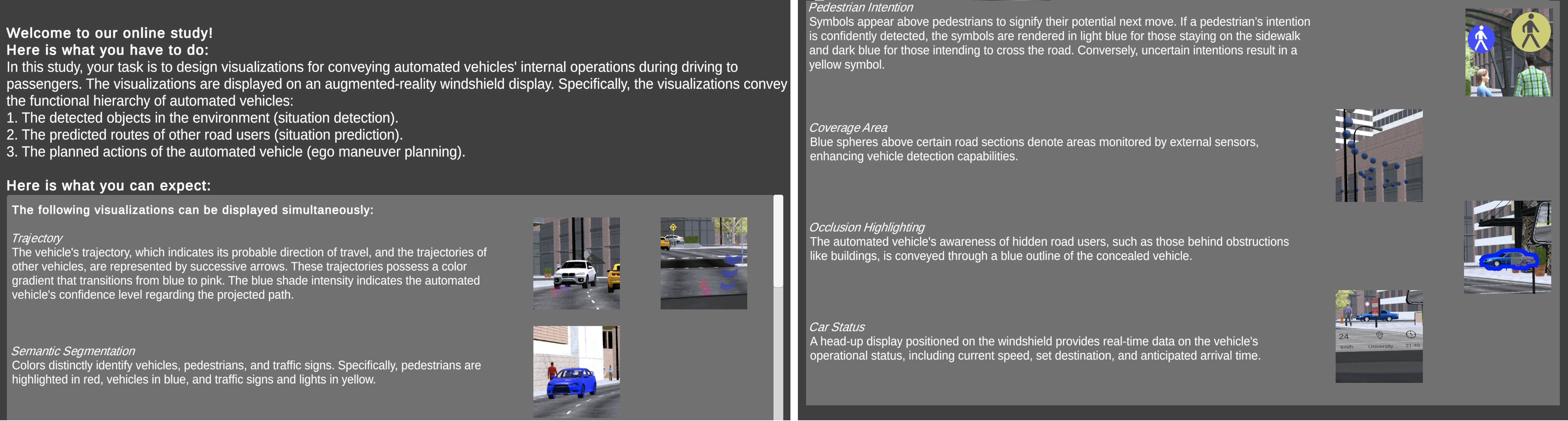}
   \caption{Excerpt of the information given to study participants at the start. Participants were also questioned about the visualizations to ensure understanding.}
   \label{fig:intro}
    \Description{The figure shows two screenshots of the initial information given to study participants. Participants were greeted with a brief introduction about the study task and short descriptions of the six visualization concepts: trajectory, pedestrian intention, semantic segmentation, occlusion highlighting, CAD-covered area, and car status.}
\end{figure*}

We included a description of the AV capabilities at the beginning of the three-day study (see below) and explained the AV functionality visualizations (see \autoref{fig:intro}).

\begin{quote}
\textit{You will see a video of a driving session in a highly automated vehicle.
The vehicle takes over lateral and longitudinal control (braking, accelerating, steering). The vehicle attempts to assess the scene and determine the intent of nearby pedestrians and cars.
While watching the video, you are supposed to imagine sitting in such an automated vehicle, follow the entire journey attentively, and then assess it.} 
\end{quote}

\begin{figure*}[ht!]
\centering
    \includegraphics[width=\linewidth]{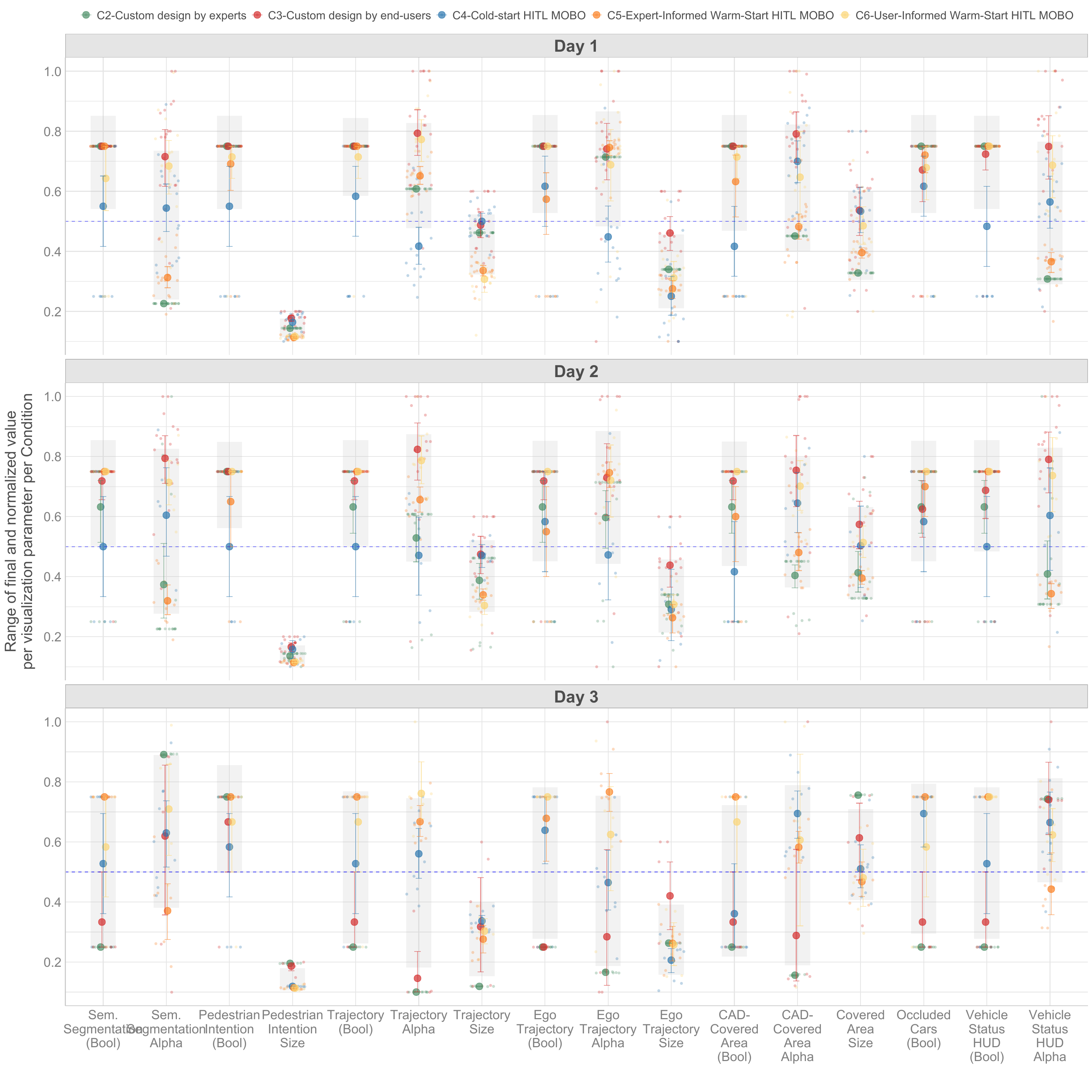}
   \caption{Final design parameter set per condition per day. The jittered \textbf{Pareto front} values per participant are presented, normalized to $[0, 1]$. Additionally, we present the \textbf{mean} parameter value per condition. The grey rectangle shows one standard deviation from the mean of all values. The x-axis shows the design parameters ordered from $p_1$ to $p_{16}$ from left to right.}
   \label{fig:final_params}
    \Description{Final design parameter set per day. Three subplots show the parameter values.}
\end{figure*}

\end{document}